\documentclass[12pt,showpacs,preprintnumbers,superscriptaddress,amsmath,amssymb,nofootinbib]{revtex4}

\usepackage{graphicx}
\usepackage{dcolumn}
\usepackage{bm}
\usepackage{amssymb}
\usepackage{amsmath}
\usepackage{epsfig}    
\usepackage{color}
\usepackage{slashed}
\usepackage{makecell}
\usepackage{float}


\begin{document}

\title{LHC Phenomenology of  Composite 2-Higgs Doublet Models}

\author{Stefania De Curtis}
\email{decurtis@fi.infn.it}
\affiliation{
INFN, Sezione di Firenze, and Department of Physics and Astronomy, University of Florence, Via
G. Sansone 1, 50019 Sesto Fiorentino, Italy}
\author{Stefano Moretti}
\email{S.Moretti@soton.ac.uk}
\affiliation{School of Physics and Astronomy, University of Southampton, Southampton, SO17 1BJ, United Kingdom}
\author{Kei Yagyu}
\email{K.Yagyu@soton.ac.uk}
\affiliation{School of Physics and Astronomy, University of Southampton, Southampton, SO17 1BJ, United Kingdom}
\author{Emine Yildirim}
\email{ey1g13@soton.ac.uk}
\affiliation{School of Physics and Astronomy, University of Southampton, Southampton, SO17 1BJ, United Kingdom}

\begin{abstract}
\noindent
We investigate the phenomenology of  Composite 2-Higgs Doublet Models (C2HDMs) of various Yukawa types 
based on the  global symmetry breaking $SO(6)\to SO(4)\times SO(2)$. 
The kinetic part and the Yukawa Lagrangian are constructed in terms of the pseudo Nambu-Goldstone Boson (pNGB) matrix and a {\bf 6}-plet of fermions under  $SO(6)$.
The scalar potential is assumed to be the same as that of the Elementary 2-Higgs Doublet Model (E2HDM) with a softly-broken discrete $Z_2$ symmetry. 
We then discuss the phenomenological differences between the E2HDM and C2HDM by focusing on
the deviations from Standard Model (SM) couplings of the discovered Higgs state ($h$) as well as on 
the production cross sections and  Branching Ratios (BRs) at the Large Hadron Collider (LHC) of extra Higgs bosons. 
We find that, even if the same deviation in the $hVV$ ($V=W,Z$) coupling is assumed in both scenarios, there appear significant differences between the E2HDM and C2HDM from the structure of the Yukawa couplings, so that production and decay features of extra Higgs bosons can be used to distinguish between the two scenarios.

\end{abstract}
\maketitle
\newpage

\section{Introduction}

After the discovery of a Higgs boson in July 2012 \cite{Aad:2012tfa,Chatrchyan:2012ufa}, an intense period of analysis
of its properties has begun and is bearing fruits. We now know that this object is very consistent with the spinless scalar
state  embedded in the SM. Following the precision measurement of its mass, around 125 GeV, its couplings to all other
states of the SM can be derived and compared with experimental data. Agreement between {the} SM and experimental results is presently
within a few tens of percent at worse, thus leaving some scope for a Beyond the SM (BSM) Higgs sector.

By bearing in mind that the discovered Higgs state has a doublet nature,
in the class of the many new physics scenarios available embedding such a structure  those among the easiest  to deal with are clearly
the 2-Higgs Doublet Models (2HDMs). Furthermore, these scenarios always include a neutral scalar Higgs state that can play the
role of the discovered one, which -- as intimated -- is very SM-like. Furthermore, they are also easily compliant with past collider data (from
LEP/SLC and Tevatron) as well as present ones (from the LHC) while still offering a wealth of new Higgs states and corresponding signals that can be searched for by the ATLAS and CMS collaborations.  In fact, a significant amount of experimental effort
at the LHC 
is presently being spared on direct searches for new Higgs bosons, in parallel with the  one of extracting their possible presence indirectly from the aforementioned precision measurements.

However, 2HDMs {\it per se} do not have the ability to solve the so-called hierarchy problem of the SM.  
An elegant way to do so though, is
to presume that the Higgs boson discovered in 2012 and its possible 2HDM companions are not fundamental particles.
 This approach is not unreasonable as any other (pseudo)scalar state found in Nature 
eventually revealed itself to be a (fermion) composite state, i.e., a mesonic state of the now standard theory
of strong interactions (QCD).  Specifically, one can construct 2HDMs in which all Higgs bosons, both neutral
and charged, both scalar or pseudoscalar, are not fundamental, rather composite. A
phenomenologically viable possibility,   wherein the mass of the lightest Higgs state is kept naturally lighter than a new strong scale (of compositeness, $f$, in the $\sim $ TeV region)   is, in particular, the one of assigning to them a pNGB nature.  {In essence, we have in mind those Composite Higgs Models (CHMs)  arising from the spontaneous symmetry breaking around the TeV scale, of the global symmetry of the strong sector \cite{Dugan:1984hq}. The resisual symmetry is explicitly broken by the SM  interactions through the \textit{partial compositeness} paradigm  \cite{Kaplan:1991dc,Contino:2006nn}.

In the minimal CHM \cite{MCHM, Contino:2006qr},  the composite version of the SM Higgs doublet, 
the only light scalar in the spectrum is indeed a pNGB (surrounded by various composite resonances, both spin-1/2 and
spin-1, generally heavier). Hence, it is natural to assume  that the new (pseudo)scalar Higgs states of a C2HDM are also pNGBs. In fact, even in the case in which they are eventually 
found to be heavier than the SM-like Higgs state, compositeness could provide a mechanism to explain  their mass differences with respect to the latter.  Finally, in the case 
of  extra Higgs doublets with no Vacuum Expectation Value
(VEV) nor couplings to quark and leptons, one could also have neutral light states as possible  composite dark matter candidates \cite{IDM}. Another example for a composite scalar dark matter candidate emerging as a pNGB is given in \cite{frig}.

C2HDMs embedding pNGBs
arising from a new strong dynamics at the TeV scale, ultimately driving Electro-Weak Symmetry Breaking (EWSB),
can be constructed  by either adopting an effective Lagrangian description (see example \cite{SILH}) invariant under {the} SM
symmetries  for  light composite $SU(2)$ Higgses or explicitly imposing a specific symmetry breaking structure containing multiple pNGBs. We take here the second approach. In detail, we will analyse 2HDMs based on  the spontaneous global symmetry breaking of an $SO(6)\to SO(4)\times SO(2)$ symmetry \cite{so6}. Within this construct, which we have
tackled in a previous paper \cite{DeCurtis:2016scv}, one can then study both the deviations of C2HDM couplings from those of a generic renormalisable E2HDM \cite{branco} as well as pursue searches for new non-SM-like Higgs signals different from the elementary case. In the $f\to \infty$ limit the  pNGB states are in fact identified with the physical Higgs states of doublet scalar fields of the E2HDM  and deviations from the E2HDM are parametrised by $\xi=v_{\rm SM}^2/f^2$, with $v_{\rm SM}$ the SM Higgs VEV.

Once the new strong  sector is integrated out, the pNGB Higgses, independently of their microscopic origin, are  described by a non-linear $\sigma$-model associated to the coset. In Ref. \cite{DeCurtis:2016scv}, we have
constructed their effective low-energy Lagrangian  according to the prescription developed by Callan, Coleman, Wess and Zumino (CCWZ) \cite{ccwz1,ccwz2}, which makes only few specific assumptions about the strong sector, namely, the global symmetries, their pattern of spontaneous breaking and the sources of explicit breaking (in our case they come from the couplings of the new strong sector with the SM fields).
The scalar potential is in the end generated by loop effects and, at the lowest order, is mainly determined by the free parameters associated to the top sector \cite{so6}. 

However, both in Ref. \cite{DeCurtis:2016scv} and here,  we will not calculate the ensuing Higgs potential \textit{a la} 
Coleman-Weinberg (CW) \cite{cw} generated by such radiative corrections, instead, we will assume the same general form  as in the E2HDM with a $Z_2$ symmetry,  {the latter imposed in order to avoid Flavor Changing Neutral 
Currents (FCNCs) at the tree level~\cite{GW}.} We do so in order to study the phenomenology of C2HDMs in a rather
model independent way, as this approach in fact allows for the most general 2HDM Higgs potential\footnote{This choice is also motivated by the fact that, in the
case in which the SM fermions are embedded in a {\bf 6}-plet representation, the leading order
terms in the perturbative and loop expansion of the potential do not provide EWSB in the composite scenario. As shown in \cite{so6}, one ought to also include next
order terms   thus generating unrelated contributions to different  operators, leading to the most general potential of the elementary version.} It is our intention to eventually construct the true version of the latter through the proper CW mechanism \cite{preparation}}. However, first we intend to infer guidance in approaching this task from the study of theoretical (i.e., perturbativity, unitarity, vacuum stability,  etc. -- the subject of Ref. \cite{DeCurtis:2016scv})
 and experimental (one of the subjects of the present paper) constraints, specifically, by highlighting the parameter space regions where differences can be found between the E2HDM and C2HDM. This will inform the choice of how to construct a phenomenologically viable and different (from the E2HDM) realisation of a C2HDM in terms of underlying gauge symmetries, their breaking patterns and the ensuing new bosonic and fermionic spectrum, that is, indeed, to settle on a specific model dependence. 

The paper is organised as follows. In Section~II we describe the C2HDM based on $SO(6)/SO(4)\times SO(2)$. 
In Section~III, the LHC phenomenology is discussed in presence of both theoretical and experimental constraints.
Conclusions are drawn in Section~IV. 
In Appendix A, relevant Feynman rules for the phenomenological study are presented.

\section{The composite two Higgs doublet model}

We construct the Lagrangian of the C2HDM based on the spontaneous breaking of the global symmetry $SO(6)\to SO(4)\times SO(2)$ at a scale $f$. 
 In this model, eight (pseudo)scalar fields emerge as pNGBs from such a breaking pattern, which constructs two
isospin doublet fields. 
In our approach, we do not specify the physics at any scale above a (large)
cutoff $\Lambda$ which is expected to be $\sim 4\pi f$ from a  na\"{\i}ve dimensional analysis~\cite{NDA}, i.e., 
we do not fix the concrete structure of the gauge  and matter contents. 
Even in this setup, the kinetic term of the pNGBs is uniquely determined by the structure of the global symmetry breaking. 
For the Yukawa sector though, we need to assume an embedding scheme for {the} SM fermions into $SO(6)$ multiplets to build the Lagrangian at  low energy. 
Although in this framework the scalar potential is generated via the CW mechanism at loop level \cite{cw}, as intimated, we assume here its renormalisable form of the E2HDM. 
This gives a sort of more general approach to the potential, namely, once the CW potential is calculated in a fixed configuration, all the potential terms can be translated into the strong sector parameters. We therefore adopt the same setup of \cite{DeCurtis:2016scv}, to which we refer the reader for further details of the
model construction.

\subsection{Two Higgs doublets as pseudo Nambu-Goldstone bosons}

We construct the $6\times 6$ pNGB matrix $U$ using the eight broken generators\footnote{We adopt the notation for the $SO(6)$ generators given in Ref.~\cite{DeCurtis:2016scv}.} 
of $SO(6)$   $T_\alpha^{\hat{a}}$ ($\alpha=1,2$ and $\hat{a}=1$-4) as 
\begin{align}
&U=\exp\left(i\frac{\Pi}{f}\right),~~\text{with}~~
\Pi\equiv \sqrt{2}\pi_{\alpha}^{\hat{a}}T^{\hat{a}}_{\alpha}=-i 
\begin{pmatrix}
0_{4\times 4} & (\pi_1^{\hat{a}},\pi_2^{\hat{a}}) \\
-(\pi_1^{\hat{a}},\pi_2^{\hat{a}})^T & 0_{2\times 2}  
\end{pmatrix}. 
\end{align}
The eight real spinless fields $\pi_\alpha^{\hat{a}}$ associated with the broken generators can be expressed through two complex doublets as
\begin{align}
\Phi_\alpha = \frac{1}{\sqrt{2}}\begin{pmatrix}
\pi_\alpha^2 + i\pi_\alpha^1 \\
\pi_\alpha^4 - i\pi_\alpha^3 
\end{pmatrix},   \label{cdoublet}
\end{align}
where the $\pi_\alpha^4$'s acquire the non-zero VEVs: $\langle \pi_\alpha^4 \rangle=v_\alpha$. 
Their ratio  is expressed as $\tan\beta = v_2/v_1$ and we define $v \equiv \sqrt{v_1^2 + v_2^2}$. 
The EW scale, $v_{\text{SM}}^{}$, related to the Fermi constant $G_F$, is expressed by $f$ and $v$ as follows:
\begin{align}
v_{\text{SM}}^{}\equiv f\sin(v/f) = (\sqrt{2}G_F)^{-1/2} \simeq 246~\text{GeV}. \label{vsm}
\end{align}

We here introduce the Higgs basis in which the physical Higgs states are separated from the NG boson states $G^\pm$ and $G^0$, which are absorbed into the 
longitudinal components of the $W^\pm$ and $Z$ bosons,  as  
\begin{align}
\begin{pmatrix}
\Phi_1 \\
\Phi_2
\end{pmatrix} 
= \begin{pmatrix}
\cos\beta & -\sin\beta \\
\sin\beta & \cos\beta 
\end{pmatrix}
\begin{pmatrix}
\Phi \\
\Psi
\end{pmatrix} , 
\end{align}
where 
\begin{align}
\Phi = \begin{pmatrix}
G^+ \\
\frac{v+h_1' + iG^0}{\sqrt{2}}
\end{pmatrix}, \quad 
\Psi = \begin{pmatrix}
H^+ \\
\frac{h_2' + iA}{\sqrt{2}} 
\end{pmatrix}. 
\end{align}
The doublet $\Psi$ contains the physical CP-odd Higgs boson $(A)$ and a pair of charged Higgs bosons $(H^\pm)$. 
As noted in Ref.~\cite{DeCurtis:2016scv}, in the Higgs basis the $G^0$, $G^\pm$ and $h_2'$ fields do not yield 
the kinetic terms in canonical form,  hence we shift these fields so as to render them to  canonical up to ${\cal O}(1/f^2)$ by
\begin{align}
G^+ \to \left(1-\frac{\xi}{3}\right)^{-1/2}G^+,\quad
G^0 \to \left(1-\frac{\xi}{3}\right)^{-1/2}G^0,\quad
h_2' \to \left(1-\frac{\xi}{3}\right)^{-1/2}h_2'. 
\end{align}
In general, the two CP-even scalar states $h_1'$ and $h_2'$ can mix with each other. Their mass eigenstates can be defined by
\begin{align}
\begin{pmatrix}
h_1' \\
h_2'
\end{pmatrix}
=
\begin{pmatrix}
\cos\theta & -\sin\theta \\
\sin\theta & \cos\theta 
\end{pmatrix}
\begin{pmatrix}
h \\
H
\end{pmatrix}, \label{theta} 
\end{align}
where $-\pi/2 < \theta \leq \pi/2$. 
We identify the mass eigenstate $h$ as the Higgs boson with a mass of 125 GeV discovered at the LHC. 

The matrix $U$ is transformed under $SO(6)$  non-linearly, i.e., $U \to gUh^{-1}$ with $g$ and $h$ being 
the transformation matrices for $SO(6)$ and $SO(4)\times SO(2)$, respectively. 
It is useful to define the linear representation of the pNGB fields from $U$ to construct the $SO(6)$ invariant Lagrangian. 
In the following, we use  the $SO(6)$ adjoint representation $\Sigma$, i.e., {\bf 15}-plet, which is reducible 
under the $SO(4)\times SO(2)$ subgroup as  
${\bf 15} = ({\bf 6},{\bf 1})\oplus  ({\bf 4},{\bf 2}) \oplus  ({\bf 1},{\bf 1})$. Namely, 
\begin{align}
\Sigma =  U\Sigma_0U^T,  \label{sigma}
\end{align}
where $\Sigma_0$ is the $SO(4)\times SO(2)$ invariant VEV parametrised as 
\begin{align}
\Sigma_0 = \begin{pmatrix}
0_{4\times 4} & 0_{4\times 2} \\
0_{2\times 4} & i\sigma_2 \label{sig2}
\end{pmatrix}. 
\end{align}
Then, the field $\Sigma$ is transformed linearly under $SO(6)$, i.e., $\Sigma \to g\Sigma g^T$. 

The kinetic terms of the eight pNGB fields can then be written in terms of $\Sigma$ as follows
\begin{align}
\mathcal{L}_{\text{kin}}=\frac{f^2}{4}\text{tr}[D_\mu \Sigma \, (D^\mu \Sigma)^T]. \label{kin15}
\end{align}
The covariant derivative $D_\mu$ is given by 
\begin{align}
D_\mu \Sigma = \partial_\mu \Sigma -i[V_\mu ,\Sigma], 
\end{align}
where
\begin{align}
V_\mu \equiv g(T_L^+ W_\mu^++T_L^- W_\mu^-) +\frac{g}{\cos\theta_W}(T_L^3-\sin^2\theta_W Q)Z_\mu +g\sin\theta_W QA_\mu, 
\end{align}
with $T_L^\pm =(T_L^1\pm i T_L^2)/\sqrt{2}$, $Q=T_L^3+T_R^3$ and $\theta_W$ being the weak mixing angle.  

In Appendix A, we give all the Feynman rules relevant to the discussion on Higgs phenomenology, which 
      are derived from the kinetic term given in Eq.~(\ref{kin15})

\subsection{Yukawa Lagrangian}

In this subsection, we construct the low-energy (below the scale $f$)  Yukawa Lagrangian.
In order to do this, we need to determine the embedding scheme of {the} SM  fermions into $SO(6)$ multiplets. 
This embedding can be justified via the mechanism based on the partial compositeness
assumption \cite{Kaplan:1991dc}, where 
elementary SM fermions mix with composite fermions in the invariant form under the SM $SU(2)_L\times U(1)_Y$ gauge symmetry but not under the global $SO(6)$ symmetry. 
Through the mixing, the $SO(6)$ invariant Yukawa Lagrangian given in terms of $\Sigma$ and composite fermions turns out to be the SM-like Yukawa Lagrangian after 
integrating out the (heavy) composite fermions. 
%

\subsubsection{Fermion embeddings}

We discuss the embeddings of {the}  SM quarks and leptons using {\bf 6}-plet representations of $SO(6)$. 
In order to reproduce the correct electric charge of {the} SM fermions, we introduce an additional 
$U(1)_X$ symmetry and assign its appropriate charge to {\bf 6}-plets. 
The electric charge $Q$ is thus given by\footnote{The $U(1)_X$ charge for the Higgs doublets $\Phi_\alpha$ must be zero to have a neutral 
component. } $Q=T_3^L+T_3^R+X$. 
In the $SO(6)$ basis, the {\bf 6}-plet fermion $\Psi_X$, with the $U(1)_X$ charge $X$  expressed as a mixture of the states in the 
$SU(2)_L\times SU(2)_R$ basis, is obtained as follows:
\begin{align}
\Psi_X =
\left[
-i\frac{\psi_{++}+\psi_{--}}{\sqrt{2}} ,
\frac{\psi_{++}-\psi_{--}}{\sqrt{{2}}}  ,
i\frac{\psi_{-+}-\psi_{+-}}{\sqrt{2}} ,
\frac{\psi_{-+}+\psi_{+-}}{\sqrt{2}}  ,
\psi_{00}     ,
\psi_{00}'    
\right]_X^T, 
\end{align}
where $\psi_{++}$, $\psi_{+-}$, $\psi_{-+}$ and $\psi_{--}$ denote 
the ($+1/2,+1/2$), ($+1/2,-1/2$), ($-1/2,+1/2$) and ($-1/2,-1/2$) state for ($T_L^3$, $T_R^3$), while
$\psi_{00}$ and $\psi_{00}'$ are singlets under $SU(2)_L$ and $SU(2)_R$, respectively. 
From this relation, we can embed
the SM quarks and leptons into the {\bf 6}-plet representation $\Psi_X$ as follows:
\begin{align}
(\Psi_{2/3})_L &\equiv  Q_L^u =
(
-id_L^{} ,
-d_L^{}  ,
-iu_L^{},
u_L^{}  ,
0 ,
0 )^T, \\
(\Psi_{-1/3})_L &\equiv Q_L^d = 
(-iu_L^{}, 
u_L^{}  ,
id_L^{} ,
d_L^{}  ,
0 ,
0 )^T, \\
(\Psi_{2/3})_R &\equiv U_R^{} =  
(0 ,
0 ,
0,
0,
0, 
u_R
)^T, \\
(\Psi_{-1/3})_R &\equiv D_R =  
(0, 
0,
0,
0,
0,
d_R)^T, \\
(\Psi_{-1})_L &\equiv L_L =
(-i\nu_L^{} ,
\nu_L^{}  ,
ie_L^{} ,
e_L^{} ,
0 ,
0 )^T, \\
(\Psi_{-1})_R &\equiv E_R^{} = 
(0, 
0,
0,
0,
0,
e_R)^T. 
\end{align} 


\subsubsection{Yukawa Lagrangian}

\begin{table}[t]
\begin{center}
{\renewcommand\arraystretch{1}
\begin{tabular}{c||ccc||ccc||ccc|ccc|ccc}\hline\hline
 &$U_R$&$D_R$&$E_R$& $(a_u,b_u)$ & $(a_d,b_d)$ & $(a_e,b_e)$  & $X_u^h$   & $X_d^h$   & $X_e^h$    & $X_u^H$    
 & $X_d^H$    & $X_e^H$   & $X_u^A$& $X_d^A$ & $X_e^A$ \\\hline 
Type-I &$-$&$-$&$-$  & $(0,\surd)$ &$(0,\surd)$ & $(0,\surd)$ & $\zeta_h$ & $\zeta_h$ & $\zeta_h$  & $\zeta_H$  & $\zeta_H$ & $\zeta_H$ & $\zeta_A$ & $\zeta_A$ & $\zeta_A$  \\\hline 
Type-II&$-$&$+$&$+$  & $(0,\surd)$ &$(\surd,0)$ & $(\surd,0)$ & $\zeta_h$ & $\xi_h$   & $\xi_h$    & $\zeta_H$  & $\xi_H$   & $\xi_H$   & $\zeta_A$ & $\xi_A$   & $\xi_A$    \\\hline 
Type-X &$-$&$-$&$+$  & $(0,\surd)$ &$(0,\surd)$ & $(\surd,0)$ & $\zeta_h$ & $\zeta_h$ & $\xi_h$    & $\zeta_H$  & $\zeta_H$ & $\xi_H$   & $\zeta_A$ & $\zeta_A$ & $\xi_A$    \\\hline 
Type-Y &$-$&$+$&$-$  & $(0,\surd)$ &$(\surd,0)$ & $(0,\surd)$ & $\zeta_h$ & $\xi_h$   & $\zeta_h$  & $\zeta_H$  & $\xi_H$   & $\zeta_H$ & $\zeta_A$ & $\xi_A$   & $\zeta_A$  \\\hline \hline 
\end{tabular}}
\caption{Charge assignment for right-handed fermions under the $C_2$ symmetry in the C2HDM. 
All the left-handed fermions $Q_L^u$, $Q_L^d$ and $L_L$ are transformed as even under $C_2$.
In the third column,  the symbol $\surd$ means non-zero $a_f$ or $b_f$.  }
\label{types}
\end{center}
\end{table}

The Yukawa Lagrangian at low energy is  given in terms of the {\bf 15}-plet of pNGB fields $\Sigma$ and the {\bf 6}-plet of fermions defined in the previous subsection:
\begin{align}
{\cal L}_Y = f\Big[\overline{Q}_L^u (a_u \Sigma - b_u \Sigma^2) U_{R} 
+\overline{Q}_L^d (a_d \Sigma - b_d \Sigma^2) D_{R}
+\overline{L}_L (a_e \Sigma - b_e \Sigma^2) E_{R}\Big] + \text{h.c.} 
\end{align}
We note that the $\Sigma^3$ term is equivalent to the $-\Sigma$ term, thus the terms with the cubic and more than cubic power of $\Sigma$
do not give any additional independent contributions to the Yukawa Lagrangian. 
The parameters $a_f$ and $b_f$ should be understood as $3\times 3$ complex matrices in  flavour space. 
This Lagrangian is rewritten, up to the order  $1/f^2$, using the complex doublet form of the Higgs fields defined in Eq.~(\ref{cdoublet}), as 
\begin{align}
{\cal L}_Y &= \sqrt{2}a_u\overline{Q}_L \left[\tilde{\Phi}_1 -\frac{1}{f^2}\tilde{\Phi}_1\left(\frac{1}{3}\Phi_1^\dagger \Phi_1 +\Phi_2^\dagger \Phi_2 \right)
+\frac{1}{3f^2}\tilde{\Phi}_2(\Phi_1^\dagger \Phi_2 +\text{h.c.})  \right]u_R^{}\notag\\ 
&+ \sqrt{2}b_u\overline{Q}_L\left[\tilde{\Phi}_2-\frac{2}{3f^2}\left(\tilde{\Phi}_1(\Phi_1^\dagger \Phi_2 + \text{h.c.})+2\tilde{\Phi}_2(\Phi_2^\dagger\Phi_2)  \right)  \right] u_R^{}\notag\\
&+\sqrt{2}a_d\overline{Q}_L \left[\Phi_1 -\frac{1}{f^2}\Phi_1\left(\frac{1}{3}\Phi_1^\dagger \Phi_1 +\Phi_2^\dagger \Phi_2 \right)
+\frac{1}{3f^2}\Phi_2(\Phi_1^\dagger \Phi_2 +\text{h.c.})  \right]d_R^{}\notag\\ 
&+ \sqrt{2}b_d\overline{Q}_L\left[\Phi_2-\frac{2}{3f^2}\left(\Phi_1(\Phi_1^\dagger \Phi_2 + \text{h.c.})+2\Phi_2(\Phi_2^\dagger\Phi_2)  \right)  \right] d_R^{}\notag\\
&+\sqrt{2}a_e\overline{L}_L\left[\Phi_1 -\frac{1}{f^2}\Phi_1\left(\frac{1}{3}\Phi_1^\dagger \Phi_1 +\Phi_2^\dagger \Phi_2 \right)
+\frac{1}{3f^2}\Phi_2(\Phi_1^\dagger \Phi_2 +\text{h.c.})  \right]e_R^{}\notag\\ 
&+ \sqrt{2}b_e\overline{L}_L\left[\Phi_2-\frac{2}{3f^2}\left(\Phi_1(\Phi_1^\dagger \Phi_2 + \text{h.c.})+2\Phi_2(\Phi_2^\dagger\Phi_2)  \right)  \right] e_R^{}, 
\label{gen_yuk}
\end{align}
where $\tilde{\Phi}_\alpha = i\sigma_2 \Phi_\alpha^*$. The  fermion mass terms, at the same order, are then extracted to be:
\begin{equation}
m_f=v_{\rm SM}\left[ a_f  c_\beta+b_f s_\beta \left( 1-\frac {\xi}{ 2} \right) \right]
\end{equation}
Clearly, the existence of two independent Yukawa matrices, $a_f$ and $b_f$, for $f=u,d,e$,  introduces  FCNCs at the tree level. As it is well known, they are induced by the fact that both doublets $\Phi_1$ and $\Phi_2$ couple to each fermion types. This 
 property is common to the E2HDM. In fact, the Yukawa Lagrangian in Eq.~(\ref{gen_yuk}),  in the limit $f\to \infty$, reproduces to so-called Type-III E2HDM. 
 
In order to avoid  FCNCs at the tree level, we impose a discrete $C_2$ symmetry~\cite{so6} as follows:
\begin{align}
&U(\pi_1^{\hat{a}},\pi_2^{\hat{a}}) \to C_2U(\pi_1^{\hat{a}},\pi_2^{\hat{a}})C_2 = U(\pi_1^{\hat{a}},-\pi_2^{\hat{a}}), 
\end{align}
where $C_2 = \text{diag}(1,1,1,1,1,-1)$. 
By this definition, $\pi_1^{\hat{a}}$ and $\pi_2^{\hat{a}}$  have 
a $C_2$-even and $C_2$-odd charge, respectively. 
Depending on the $C_2$ charge assignment of the right-handed fermions, we can define four independent types of Yukawa interactions, just like 
the  softly-broken $Z_2$ symmetric version of the E2HDM \cite{Barger,Grossman,typeX}, as shown in Tab.~\ref{types}. For example,
the Type-I Yukawa interaction is obtained by taking $a_f=0$. 

In the $C_2$ symmetric case, we obtain the following interaction terms in the mass eigenbasis of the fermions:
\begin{align}
{\cal L}_Y 
&=\sum_{f=u,d,e}\frac{m_f}{v_{\text{SM}}}\bar{f}\Big(X_f^h h +X_f^H H - 2iI_f X_f^A  \gamma_5 A \Big)f\notag\\
&+ \frac{\sqrt{2}}{v_{\text{SM}}}\bar{u}V_{ud}(m_dX_d^A\,P_R-m_uX_u^A\,P_L  )d\,H^+  + \frac{\sqrt{2}}{v_{\text{SM}}}\bar{\nu}m_eX_e^AP_R\, e \, H^+ +\text{h.c.}  , 
\end{align}
where $I_f=+1/2~(-1/2)$ for $f=u~(d,e)$, $V_{ud}$ are the CKM matrix elements and $P_{L,R}$ are the projection operators for left and right handed fermions.
The coefficients $X_f^{h,H,A}$ can be either $\zeta_{h,H,A}$ or $\xi_{h,H,A}$ as shown in Tab.~\ref{types}, and their 
expressions, at the first order in $ \xi $, are given by: 
\begin{align}
\zeta_h & =  \left(1-\frac{3}{2}\xi \right)c_\theta + s_\theta \cot\beta, \quad \xi_h  =  \left(1-\frac{\xi}{2} \right)c_\theta  -s_\theta  \tan\beta, \label{xihh}\\
\zeta_H & =  -\left(1-\frac{3}{2}\xi \right)s_\theta + c_\theta \cot\beta, \quad \xi_H  =  -\left(1-\frac{\xi}{2} \right)s_\theta  - c_\theta  \tan\beta, \label{xibh} \\
\zeta_A & =  \left(1+\frac{\xi}{2} \right)\cot\beta, \quad 
\xi_A    =  -\left(1-\frac{\xi}{2} \right)\tan\beta. 
\end{align}
In the limit of $\xi\to 0$, 
these coefficients get the same form as the corresponding ones in a softly-broken $Z_2$ symmetric version of the E2HDM~\cite{typeX}.


\subsection{Potential}

We adopt the same form of the potential as in the E2HDM. We  have in total eight parameters, 
which can be translated into eight physical inputs, as explicitly done in Ref.~\cite{DeCurtis:2016scv}:
\begin{align}
m_h,~~m_H^{},~~m_A^{},~~m_{H^\pm}^{},~~M^2,~~v_{\text{SM}}^{},~~\tan\beta~~\text{and}~~s_\theta, \label{phys}
\end{align}
where $m_h,~~m_H^{},~~m_A^{}$ and $m_{H^\pm}^{}$ are respectively the mass of $h$, $H$, $A$ and $H^\pm$, among which 
$m_h$ should be fixed to be 125 GeV. The parameter $M$ describes the soft breaking scale of the $C_2$ symmetry. 

\section{Phenomenology}

In this section, we discuss how we can discriminate the C2HDM from the E2HDM. 
Firstly, we  discuss the constraints on the parameter space in the C2HDM from various (null) searches of extra Higgs bosons at collider experiments (Sec.~III-A). 
Secondly, we focus on deviations in the SM-like Higgs boson ($h$) couplings from the SM predictions (Sec.~III-B). 
Thirdly, we discuss the difference in the properties of the decay (Sec.~III-C) and production mechanisms at the LHC (Sec.~III-D)
of the extra Higgs bosons.

{Before proceeding further though, we ought to note now that CHMs  usually also predict  heavy spin-1 and spin-1/2 states  via the partial compositeness mechanism. These extra particles can potentially enter the ensuing phenomenological analysis through  loop induced couplings of the  composite  Higgs bosons to $Z$, photons and gluons. Contributions from  these  particles   to  loop-induced Higgs production and/or decay modes were studied in detail  in Ref.~\cite{4DCHM} (albeit for a single doublet realisation of a CHM). It was  found therein that the individual extra gauge boson contributions to the loop induced couplings of the SM-like $h$ state are always negligible while  this is not the case for the extra fermion ones. However, the  extra fermion effects are not dramatically large, i.e., at the $ \mathcal{O} $(1\%) level in the $h\rightarrow \gamma \gamma,\gamma Z$  cases and   $\mathcal{O}$(10\%) in the $h\to gg$ case. Thus, Ref.~\cite{4DCHM} indicates that all such contributions to $h$ phenomenology at the LHC are not crucial for our purposes. Similarly, we expect that these  extra resonances (both bosons and fermions)  will not affect significantly the $H, A \rightarrow gg,\gamma\gamma$ and $Z\gamma$ partial widths either. Hence, in our forthcoming C2HDM analysis at the LHC, we will not include all such effects.} 

\subsection{Constraints from collider experiments}

We start discussing constraints on the parameter space of the E2HDM and C2HDM  
from data collected at LEP, Tevatron and  LHC by using the {\tt HiggsBounds}~\cite{HB1,HB2,HB3,HB4} (v4.3.1) package.
This tells us if a given set of model parameters is allowed at 95\% Confidence Level (CL)  by various (null) searches of Higgs bosons.

\begin{figure}[t]
\begin{center}
\includegraphics[width=0.21\linewidth,angle=270]{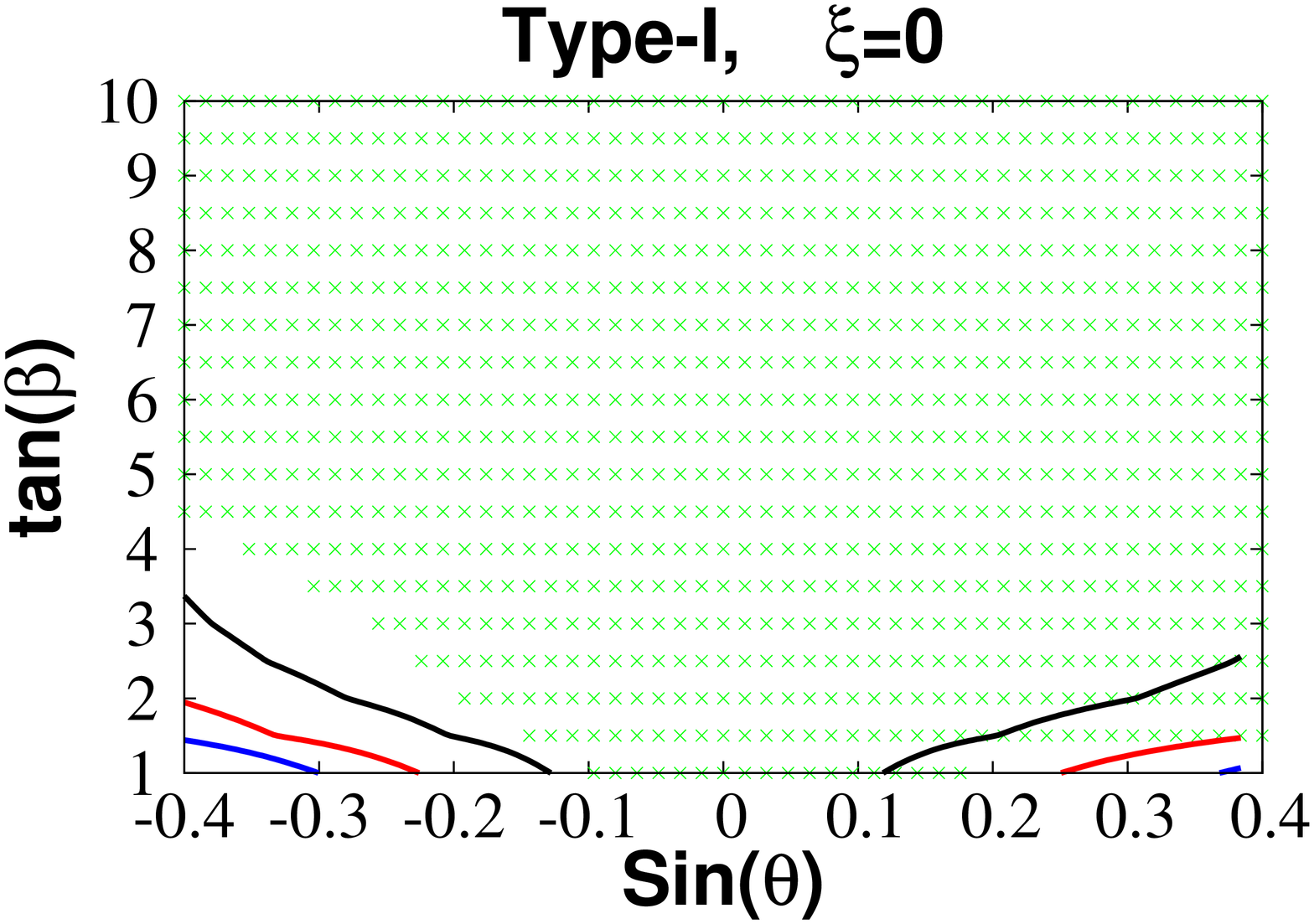}\hspace{4.mm}
\includegraphics[width=0.21\linewidth,angle=270]{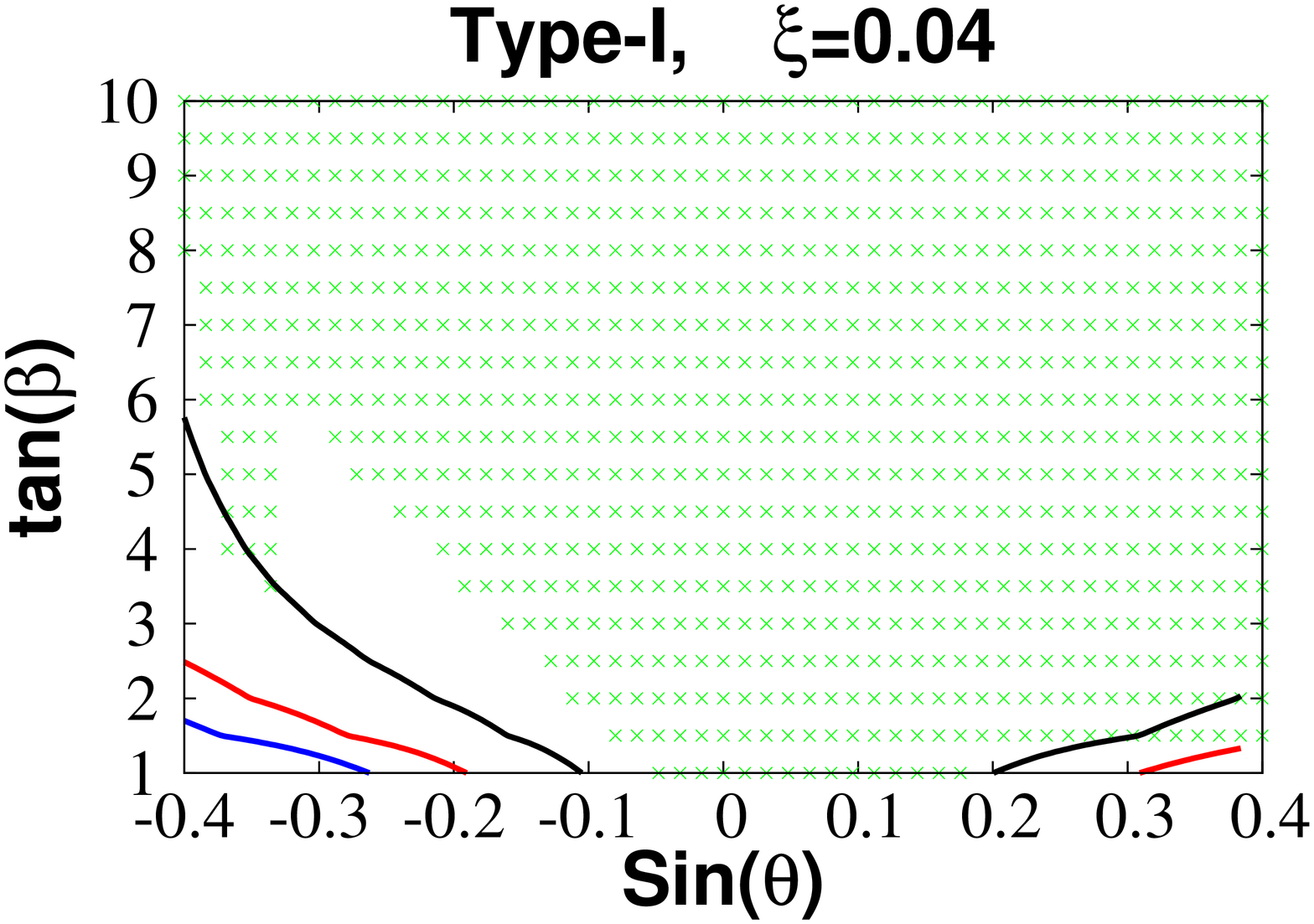}\hspace{3.25mm}
\includegraphics[width=0.21\linewidth,angle=270]{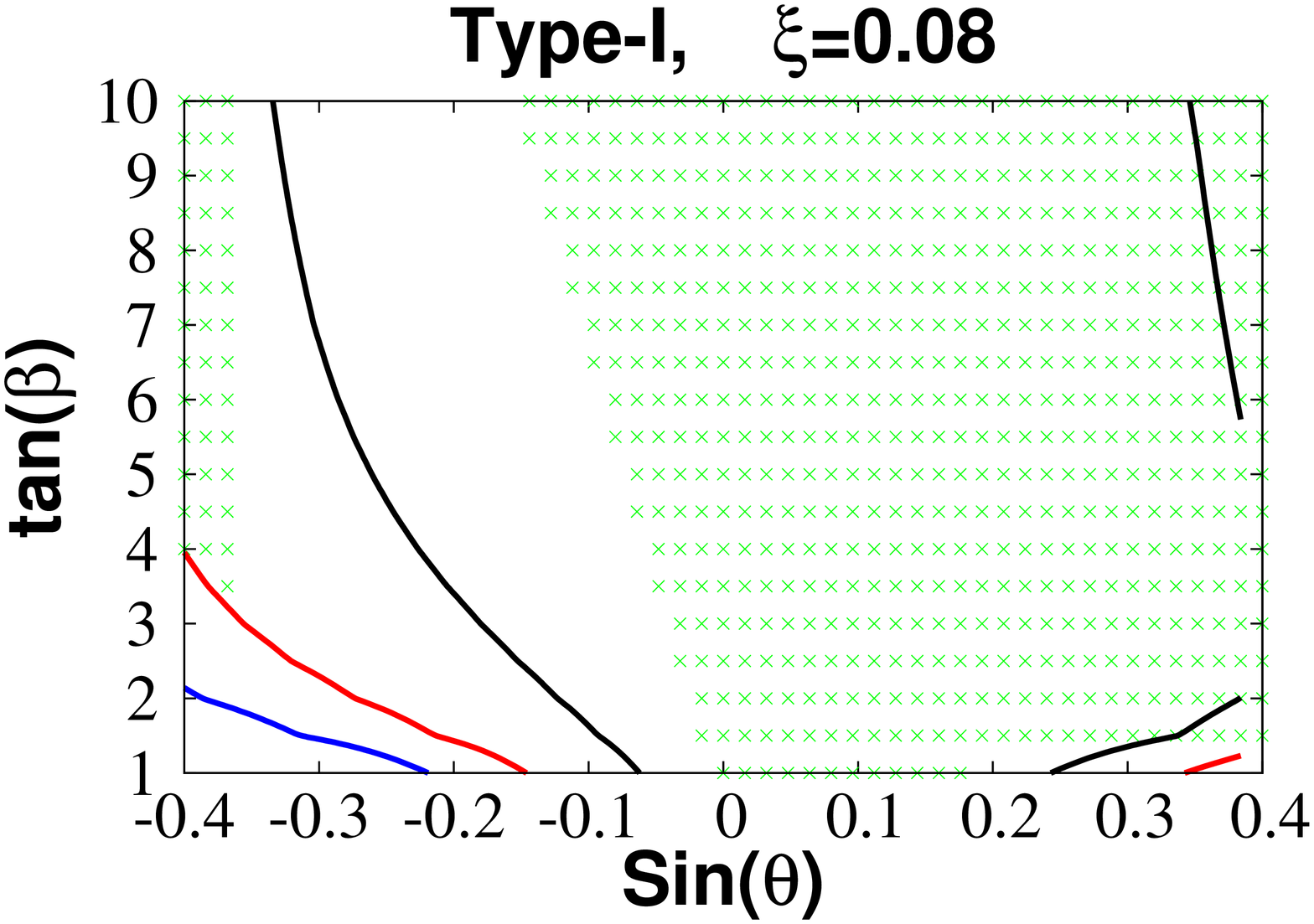}\hspace{3.25mm}\vspace{5mm}

\includegraphics[width=0.21\linewidth,angle=270]{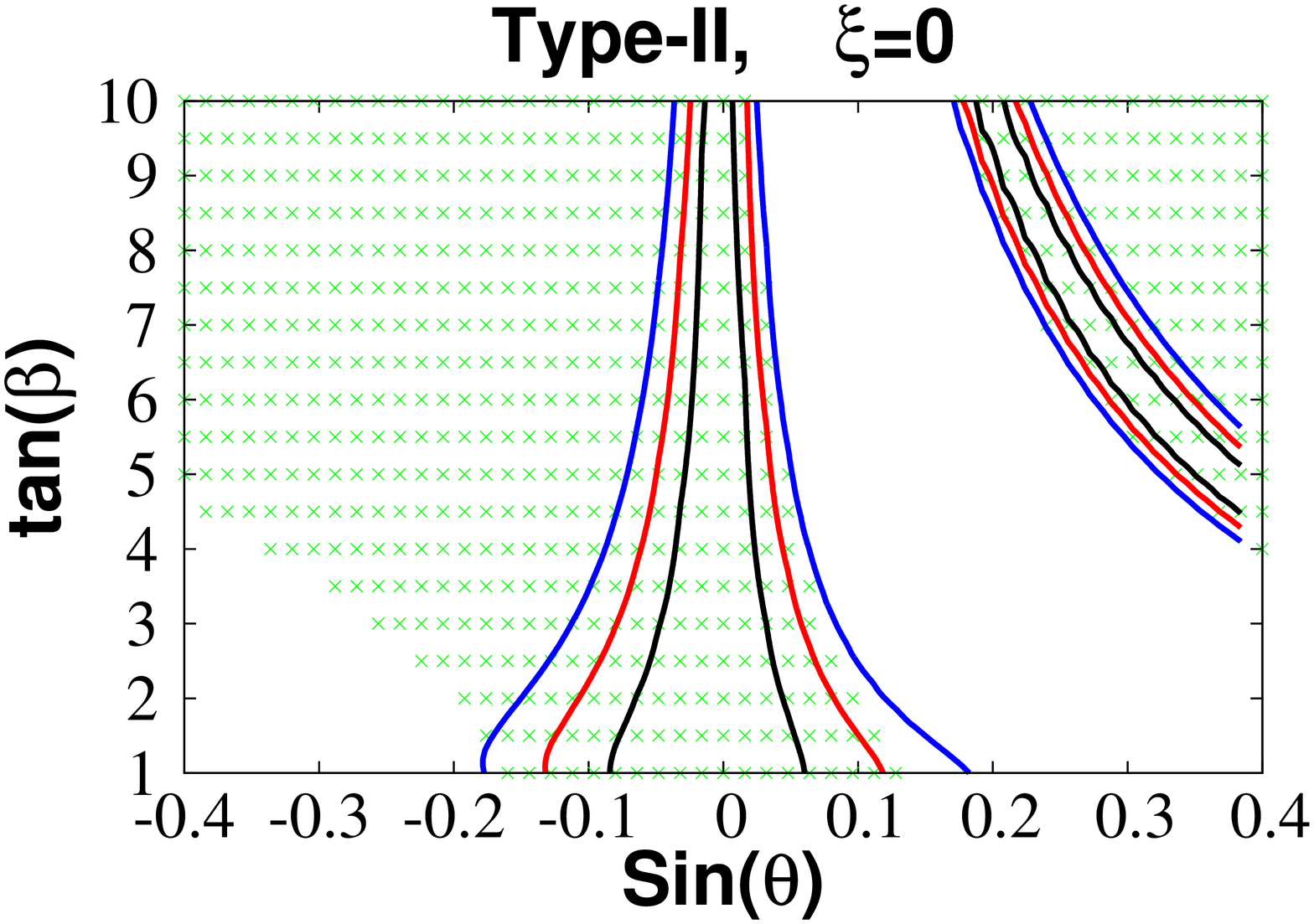}\hspace{4.mm}
\includegraphics[width=0.21\linewidth,angle=270]{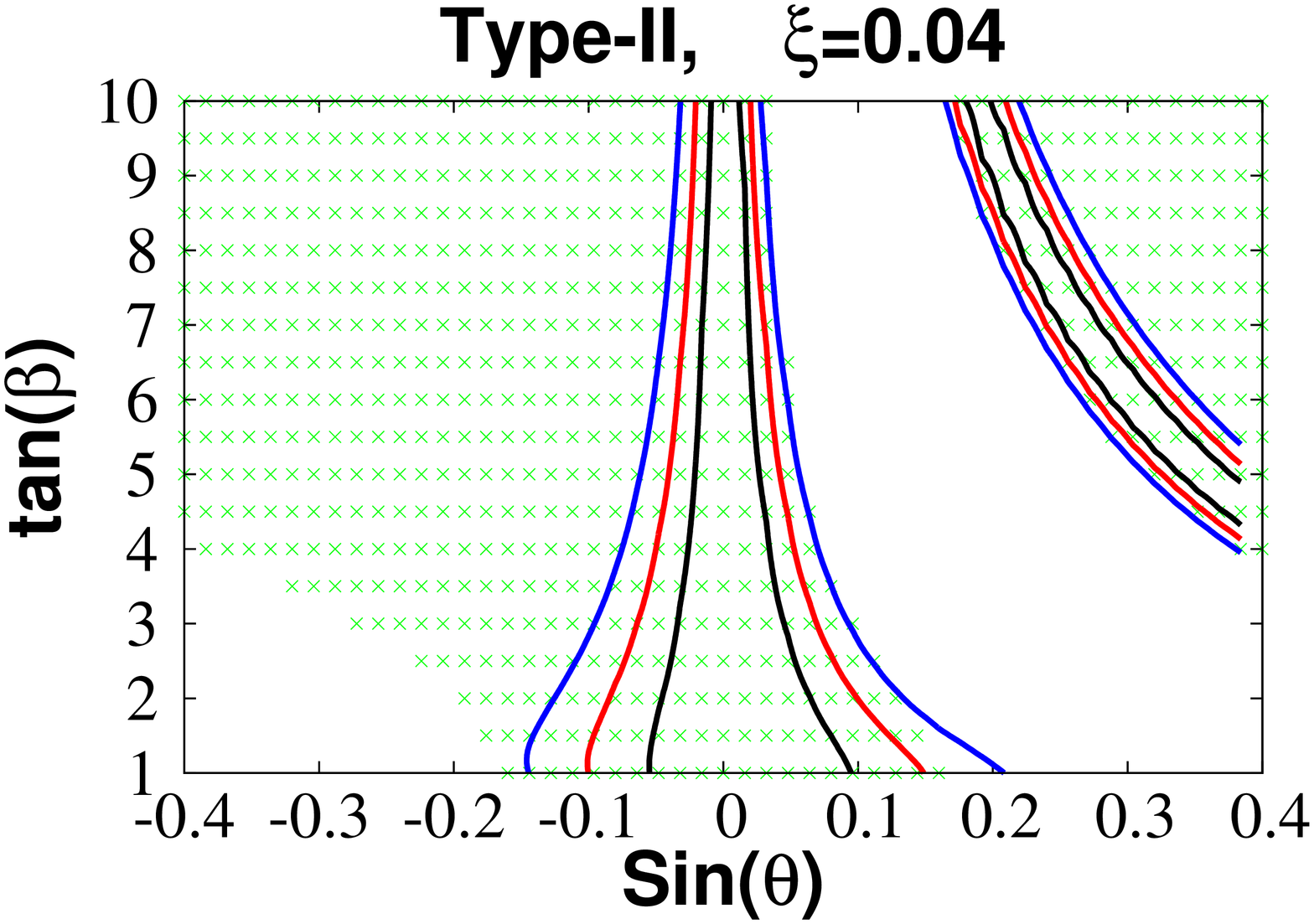}\hspace{3.25mm}
\includegraphics[width=0.21\linewidth,angle=270]{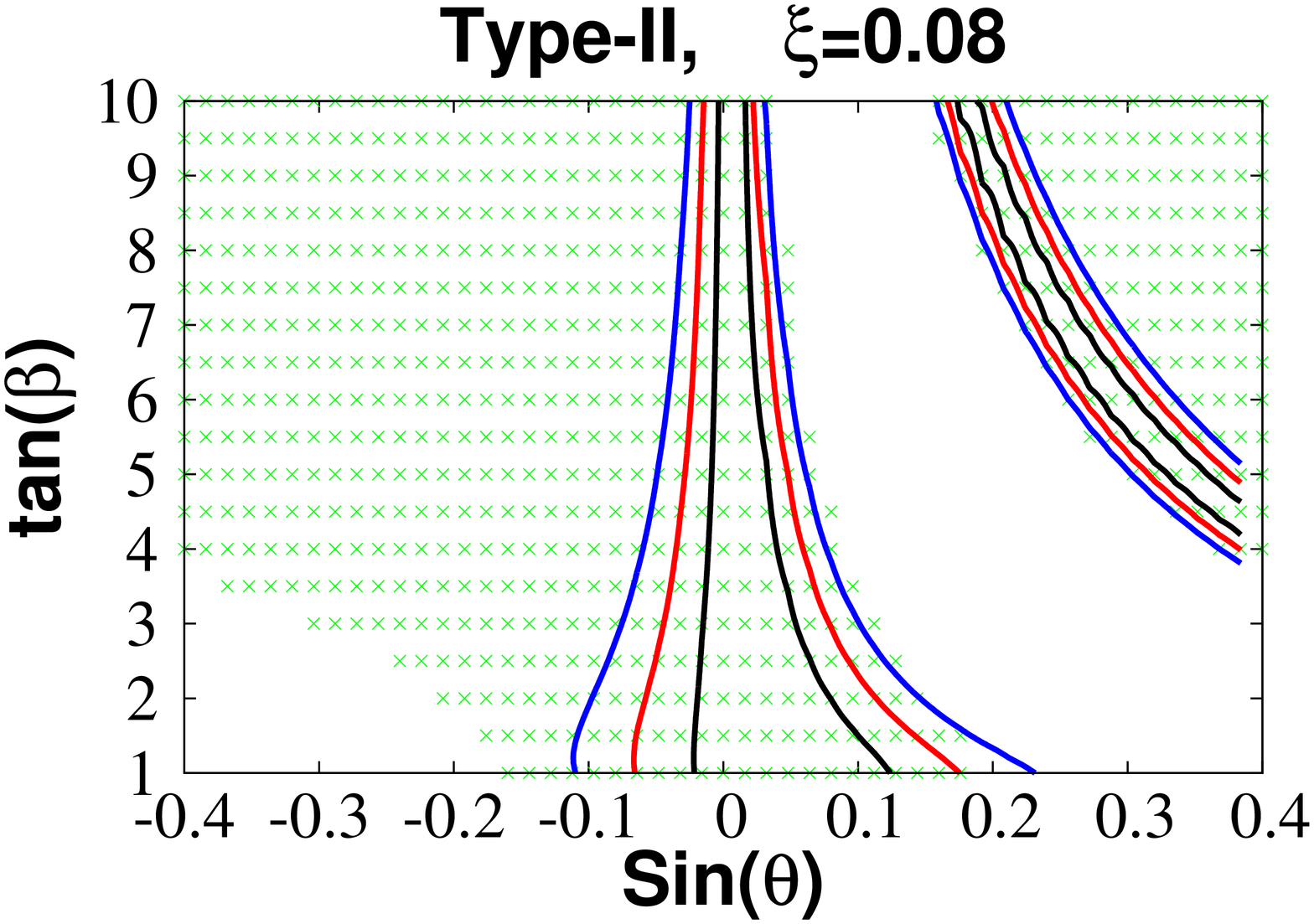}\hspace{3.25mm}\vspace{5mm}

\includegraphics[width=0.21\linewidth,angle=270]{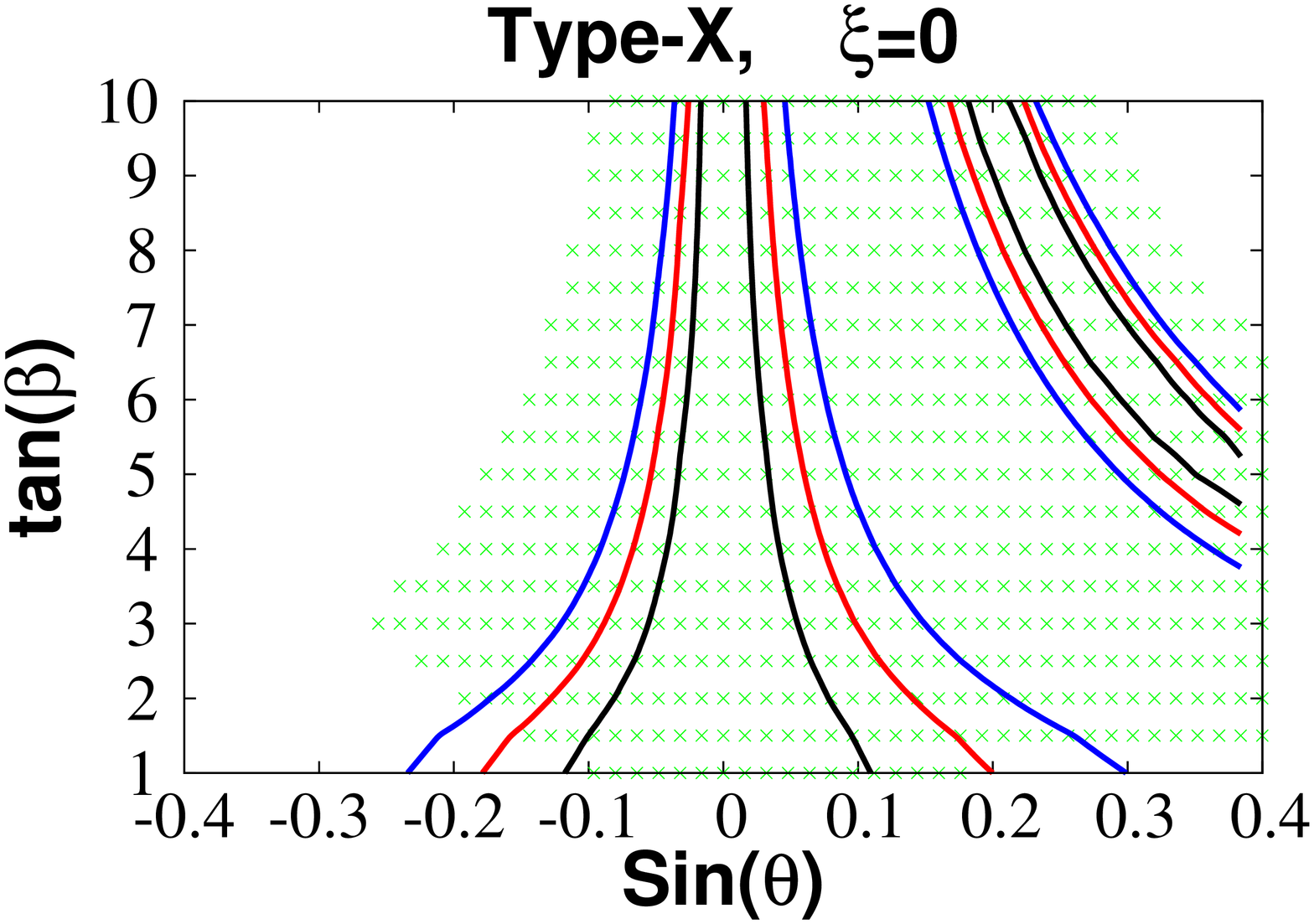}\hspace{4.mm}
\includegraphics[width=0.21\linewidth,angle=270]{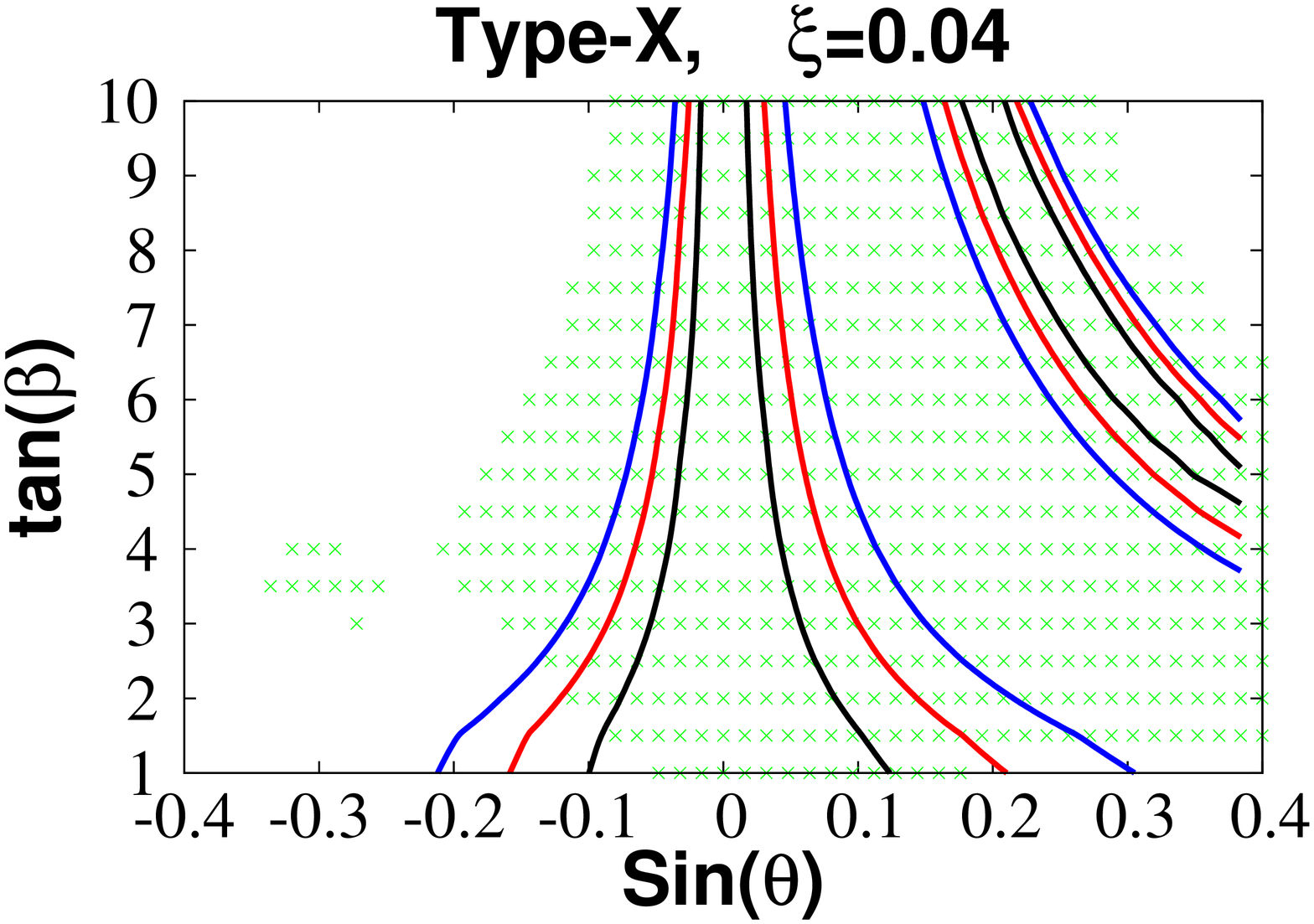}\hspace{3.25mm}
\includegraphics[width=0.21\linewidth,angle=270]{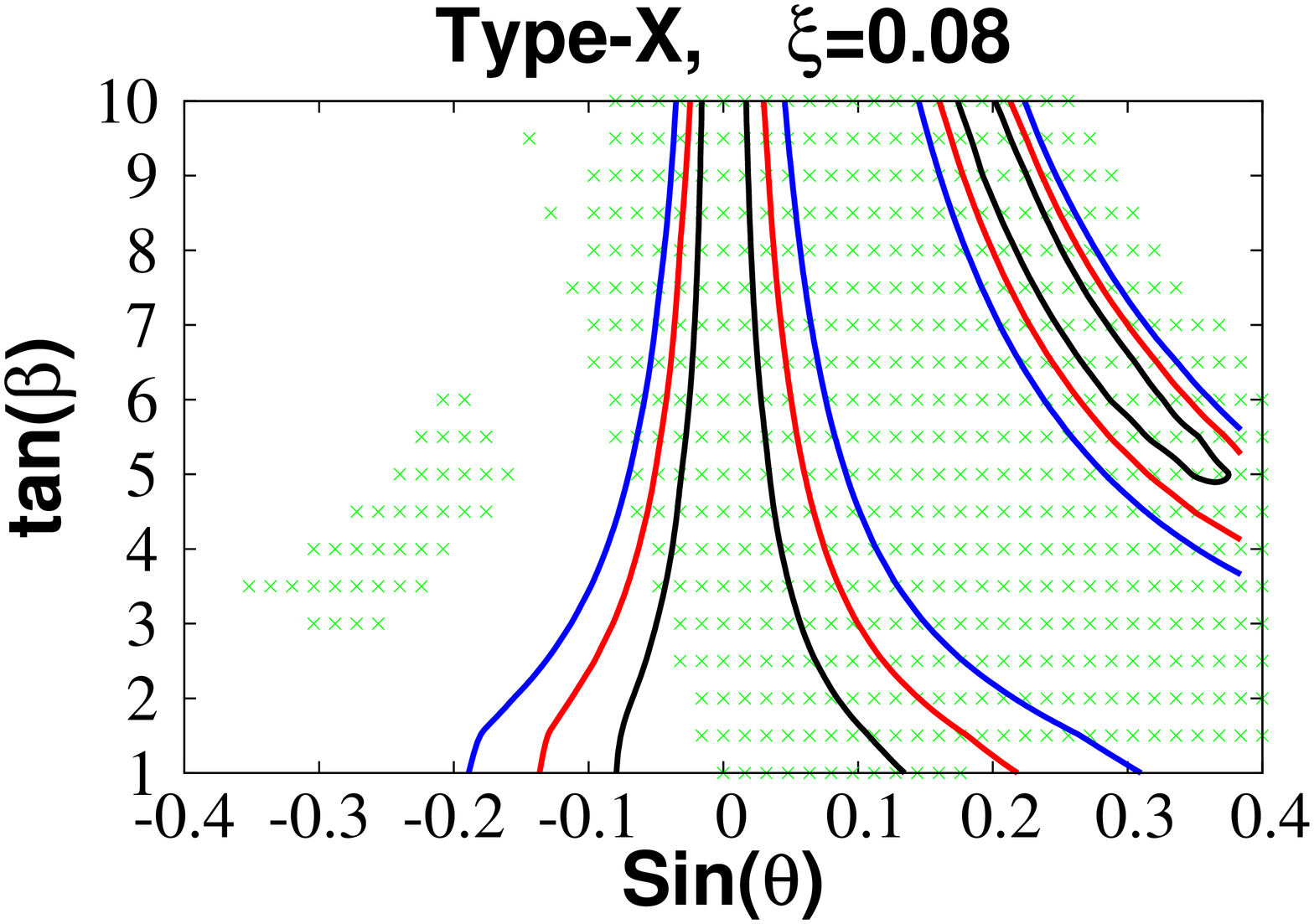}\hspace{3.25mm}\vspace{5mm}

\includegraphics[width=0.21\linewidth,angle=270]{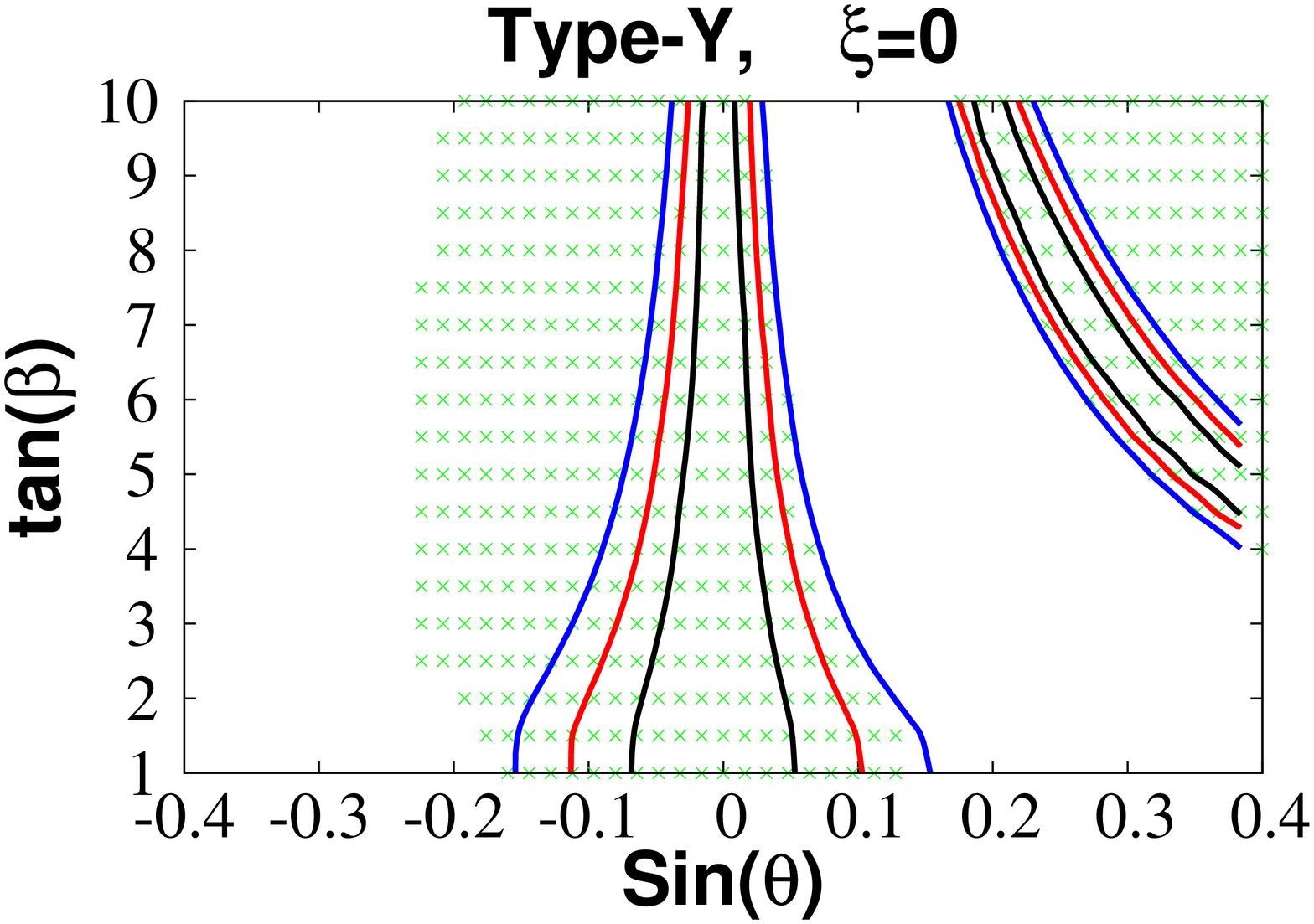}\hspace{4.mm}
\includegraphics[width=0.21\linewidth,angle=270]{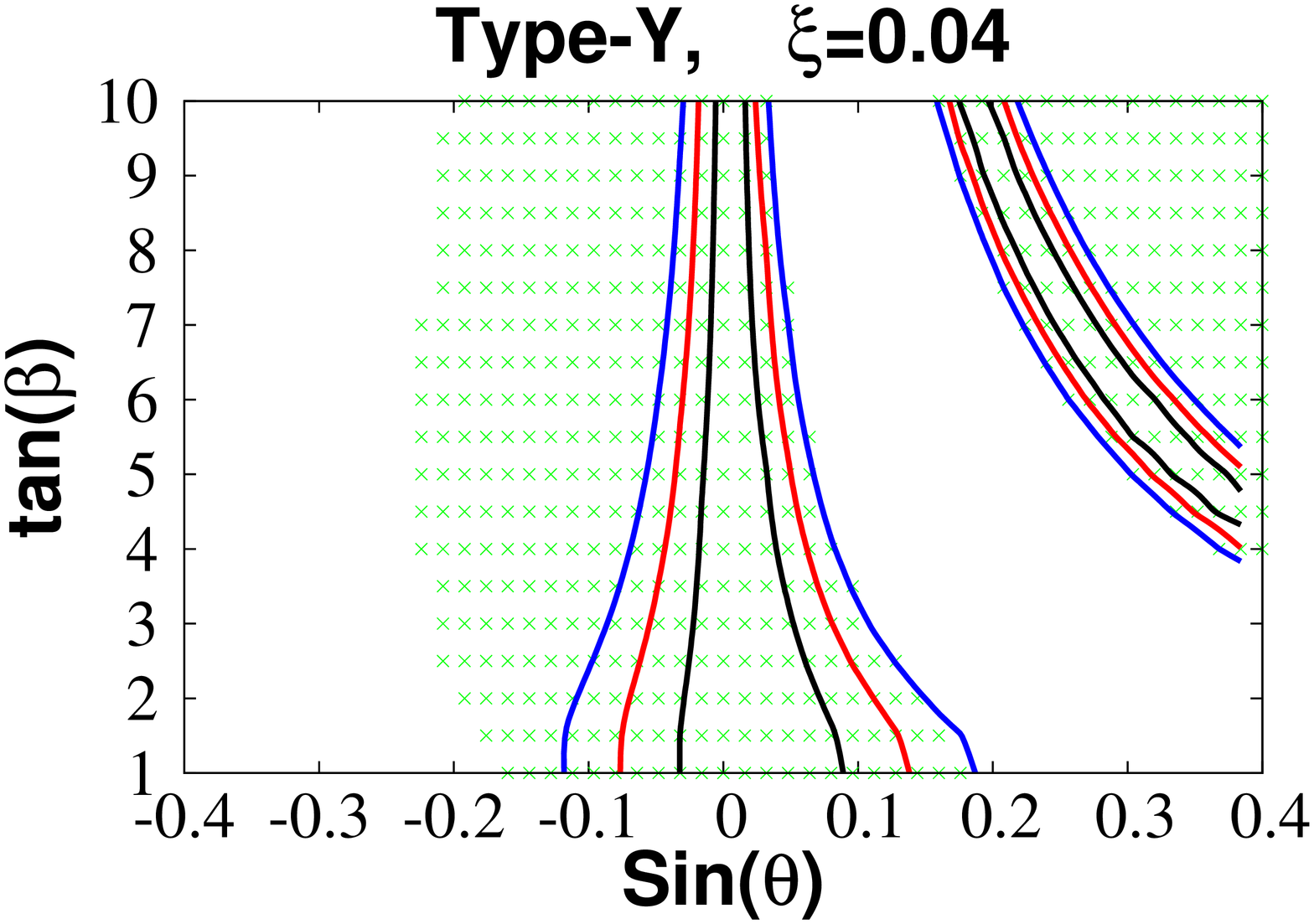}\hspace{3.25mm}
\includegraphics[width=0.21\linewidth,angle=270]{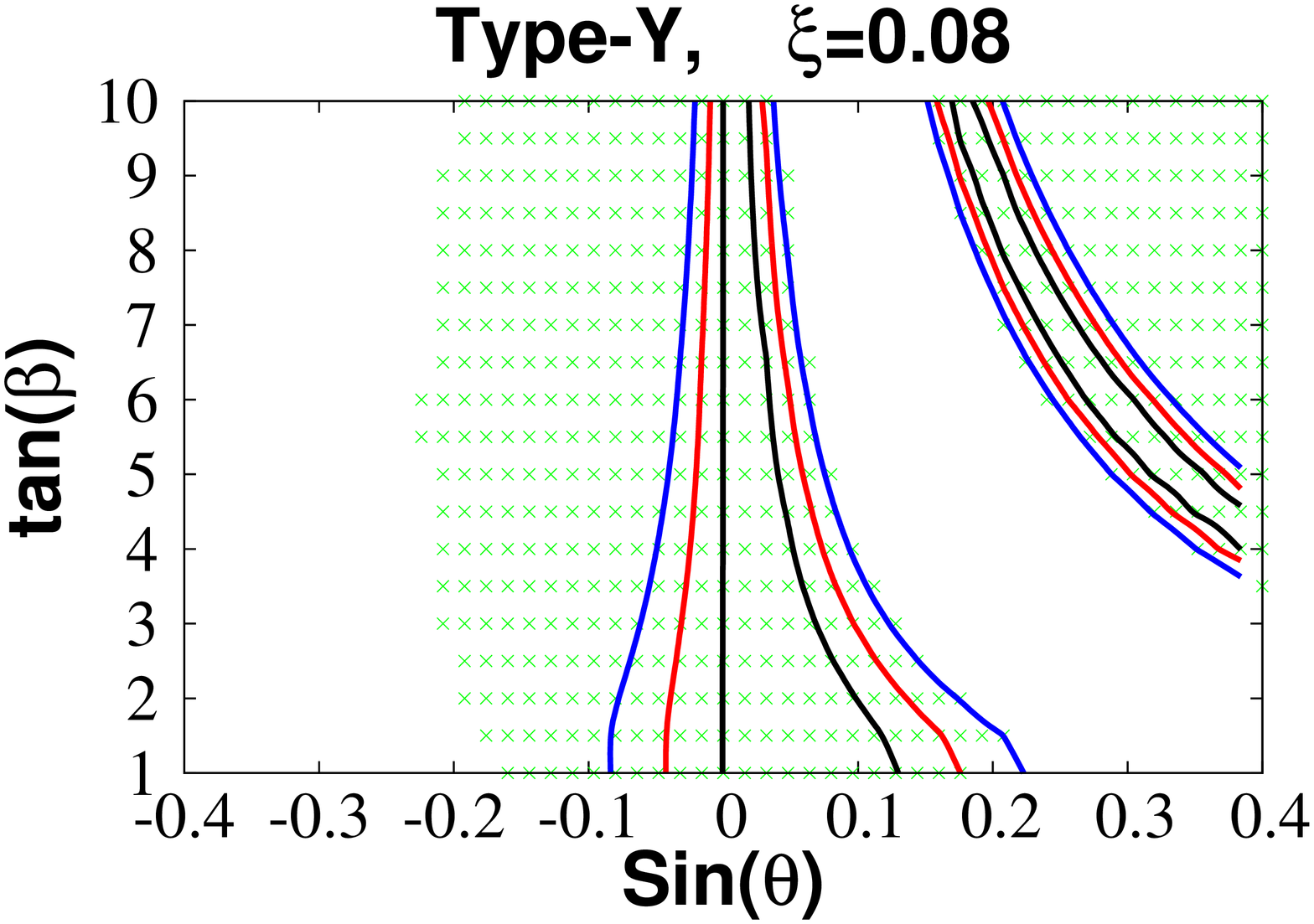}\hspace{3.25mm}

\caption{Regions  allowed at 95\% CL from LEP, Tevatron and LHC experiments in the Type-I, -II, -X and -Y  C2HDMs (green shaded). The black, red and blue curves display the contours for $ \Delta \chi^{2}=$ 2.30 (68.27\% CL), 6.18 (95.45\% CL) and 11.83 (99.73\% CL), respectively. As for the reference input values, we take $m_h$ = 125 GeV,  $m_H = m_{H^{\pm}} = m_A$ = 500 GeV and $M = 0.8~m_A$.
The first, second, and third column of panels  show the results with $ \xi=~ 0,~ 0.04,~0.08$, respectively.
}
\label{HB}
\end{center}
\end{figure}

In Fig.~\ref{HB}, we show the allowed regions (green shaded) in the $(\sin\theta,\tan\beta)$ plane using {\tt HiggsBounds}. 
We can see that larger values of $\xi$ give more excluded regions in the Type-I C2HDM, but in the other three models
the $\xi$ dependence is not so significant.  
In particular, in the Type-I C2HDM with $\xi=0.08$ 
negative values of $\sin\theta$ are mostly ruled out mainly because of the positive deviation of the signal strength for the 
vector boson fusion production of $h$ (the SM-like state) decaying into $W^+W^-$~\cite{cms3} as compared to the Type-I E2HDM (corresponding to $\xi=0$ and shown in the first column  of Fig.~\ref{HB}). 
%
%
In the Type-X C2HDM, additional exclusion parameter regions appear for larger  $\xi$, which is   mostly due to the same reason as in the Type-I C2DHM. 
In the Type-II and -Y C2HDMs, the excluded regions almost do not depend upon the $\xi$ value. In Tab.~\ref{channels}  we list the Higgs search channels most responsible for the exclusions.

\begin{table}[t]
\resizebox{\columnwidth}{!}{
\scalebox{0.9}{
\begin{tabular}{ll}
\begin{tabular}{|c|c|c|c|c|c|}\hline
Model Type &  $ \xi=0.08 $ &$ \xi=0.04 $& $ \xi=0 $\\ \hline
Type I & \makecell{$ pp \to  H \to  ZZ \to  4\ell $ \cite{atlas1}\\$qQ \to q'Q'h \to WW \to 2\ell 2\nu$ \cite{cms3}}& \makecell{$ pp \to  H \to  ZZ \to  4\ell $ \cite{atlas1}\\$qQ \to q'Q'h \to WW \to 2\ell 2\nu$ \cite{cms3}}&\makecell{$ pp \to  H \to  ZZ \to  4\ell $ \cite{atlas1}\\$qQ \to q'Q'h \to WW \to 2\ell 2\nu$ \cite{cms3}} \\ \hline

Type II    &  \makecell{$ pp \to  H \to  ZZ \to  4\ell $ \cite{atlas1}\\  $ pp \to  h \to  ZZ \to  4\ell $ \cite{cms4}} & \makecell{$ pp \to  H \to  ZZ \to  4\ell $ \cite{atlas1}\\ $qQ \to q'Q'h \to WW \to 2\ell 2\nu$ \cite{cms3}\\  $ pp \to  h \to  ZZ \to  4\ell $ \cite{cms4}} &\makecell{$ pp \to  H \to  ZZ \to  4\ell $ \cite{atlas1}\\$qQ \to q'Q'h \to WW \to 2\ell 2\nu$ \cite{cms3}\\  $ pp \to  h \to  ZZ \to  4\ell $ \cite{cms4}\\ $ pp \to  h \to  WW^{\ast} \to  \ell\nu\ell\nu $ \cite{atlas2}}\\ \hline

 Type X   & \makecell{$ pp \to  H \to  ZZ \to  4\ell $ \cite{atlas1}\\$qQ \to q'Q'h \to WW \to 2\ell 2\nu$ \cite{cms3}\\ gg $  \to  \phi (h,H) \to  \tau \tau $ \cite{cms5} \\ $ pp \to  h \to  \tau \tau $ \cite{cms6}\\ $ p p \to  V h \to  V \tau \tau $ \cite{cms7}}&\makecell{$ pp \to  H \to  ZZ \to  4\ell $ \cite{atlas1}\\$qQ \to q'Q'h \to WW \to 2\ell 2\nu$\cite{cms3}\\ gg $ \to  \phi (h,H) \to  \tau \tau $ \cite{cms5} \\ $ pp \to  h \to  \tau \tau $ \cite{cms6}\\ $ p p \to  V h \to  V \tau \tau $ \cite{cms7}}&\makecell{$ pp \to  H \to  ZZ \to  4\ell $ \cite{atlas1}\\ $ pp \to  h \to  \tau \tau $ \cite{cms6}}\\  \hline

Type Y &\makecell{$ pp \to  H \to  ZZ \to  4\ell $ \cite{atlas1}\\ $ pp \to  H \to  hh \to  4b $ \cite{cms1}\\ $ pp \to  h \to  ZZ \to  4\ell $ \cite{cms4}}&\makecell{$ pp \to  H \to  ZZ \to  4\ell $ \cite{atlas1}\\ $ pp \to  H \to  hh \to  4b $ \cite{cms1}\\ $ pp \to  h \to  ZZ \to  4\ell $ \cite{cms4}}& \makecell{  $ pp \to  H \to  ZZ \to  4\ell $ \cite{atlas1}\\ $ pp \to  H \to  hh \to  4b $ \cite{cms1}\\ $ pp \to  h \to  ZZ \to  4\ell $ \cite{cms4}\\ $ pp \to  h \to  WW^{\ast} \to  \ell\nu\ell\nu $ \cite{atlas2}} \\  \hline
\end{tabular}
\end{tabular}
}}
\vspace{5mm}
\caption{Higgs search channels most responsible for excluding parameter points  in Fig.~\ref{HB}. }
\label{channels}
\end{table}

In Fig.~\ref{HB} we also present the compatibility of the signal strengths of the SM-like Higgs boson $h$ based on a $\Delta\chi^2$ analysis by using the 
{\tt HiggsSignals}~\cite{HS} (v1.4.0) package. 
In this figure, the black, red  and blue contours  respectively show the compatibility with 68.27\% CL, 95.45\% CL and 99.73\% CL from the minimum value of 
$\chi^2$ in the $(\sin\theta,\tan\beta)$ plane. Apart from the Type-I case, which reveals a better compliance with
the LHC data (this is after all the scenario which more closely resembles the SM), the other three types respond similarly
to the LHC Higgs data. Overall, the $\xi$ dependence is only marginally evident, being more pronounced for Type-I.

\begin{figure}[t]
\begin{center}
\includegraphics[width=80mm]{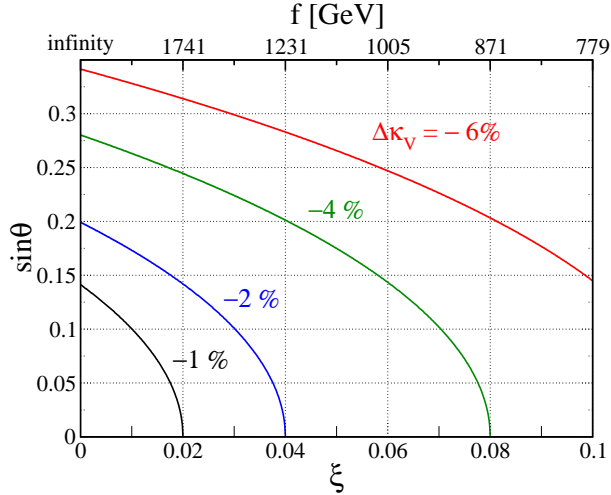}
\caption{Contour plot for the deviation in the $hVV$ couplings $\Delta\kappa_V^{}=\kappa_V^{}-1$ from the SM prediction. 
}
\label{kv}
\end{center}
\end{figure}

\subsection{Deviation in the Higgs boson couplings}

In both the E2HDM and the C2HDM, the Higgs boson couplings can deviate from the SM predictions. 
However, the pattern of deviations can be different between these two scenarios. 
In order to discuss these, it is convenient to define the scaling factor 
$\kappa_X^{}$ for the $hXX$ couplings by $\kappa_X^{} = g_{hXX}^{\text{NP}}/g_{hXX}^{\text{SM}}$  and $\Delta \kappa_X=\kappa_X-1$. 
In the C2HDM, these deviations are given at the tree level by 
\begin{align}
\kappa_V^{} = \left(1-\frac{\xi}{2} \right)c_\theta~~(V=W,Z),\quad 
\kappa_f^{} = X_f^h = \zeta_h\text{~~or~~}\xi_h~~(f=u,d,e). 
\end{align}
Those for the E2HDM can be easily obtained by taking $\xi\to 0$ corresponding to $f\to\infty$. 
We can see that there are two sources giving $\kappa_X^{} \neq 1$ in the C2HDM, i.e., 
non-zero values of $\xi$ and $\theta$. 
Conversely, only $\theta \neq 0$ gives $\kappa_X^{}\neq 1$ in the E2HDM\footnote{Radiative corrections can also modify the $hVV$ couplings, 
but their typical magnitude is less than 1\%~\cite{thdm_rad} with respect to the tree level prediction in the E2HDM. }. 
Therefore, for a given measured value of $\kappa_X^{}$, the value of $\theta$ is determined in the E2HDM while only
the combination $(\theta,\xi)$ is determined in the C2HDM. 

In Fig.~\ref{kv}, we plot the contour for  $\Delta\kappa_V^{}$ as a function of $\xi$ and $\sin\theta$. We note that 
there is no sign dependence of $\sin\theta$ in this plot. 
From this figure, it is clear that a fixed value of the deviation $\Delta\kappa_V^{}$ at $\xi=0$ which corresponds to  the E2HDM
can be reproduced in the C2HDM by different parameter  with non-zero $\theta$ and/or $\xi$. 
For example, the deviation $|\Delta\kappa_V^{}|=2\%$ can be reproduced by e.g., $(\xi,\theta)=(0.04,0)$, $(0.03,0.1)$ and $(0,0.2)$. 
This result suggests an interesting consequence, namely, even if there is no mixing between the CP-even Higgs bosons $h$ and $H$, in the C2HDM, 
we can have a non-zero deviation in the $hVV$ couplings. As a result, we will find a  significant difference in the two scenarios for the decay 
Branching Ratios (BRs) of the extra Higgs bosons
for a given value of $\Delta\kappa_V^{}$, which will be discussed in the succeeding subsections. 

\begin{figure}[t]
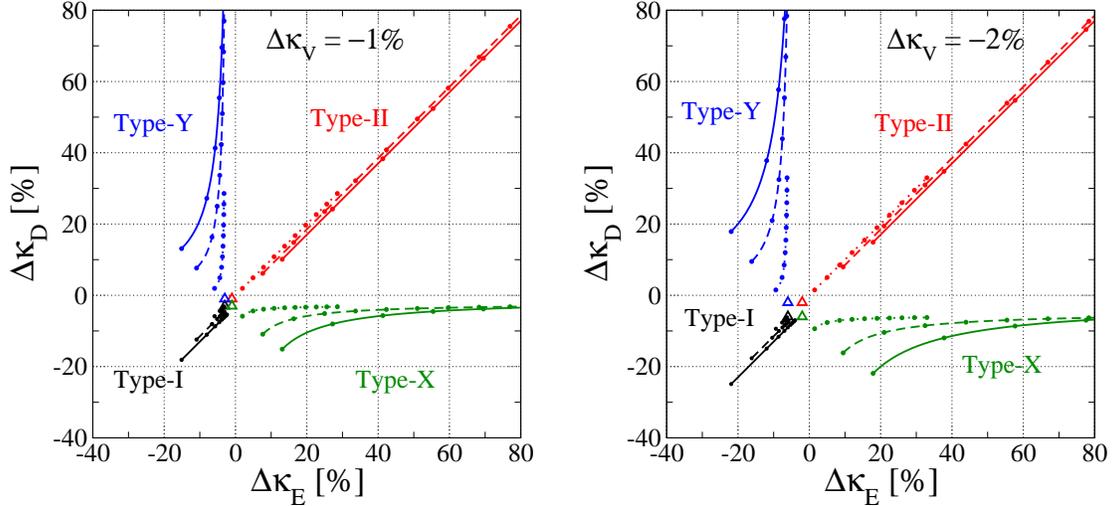

\begin{center}
\includegraphics[width=70mm]{finger_kv1.eps}\hspace{5mm}
\includegraphics[width=70mm]{finger_kv2.eps}
\caption{Deviations in the Yukawa couplings 
on the $\Delta \kappa_E^{}$ ($E$ stands for a charged lepton) 
and $\Delta \kappa_D^{}$ ($D$ stands for a down-type quark) plane in the C2HDMs with $s_\theta < 0$. 
The left (right) panel shows the case for $\Delta\kappa_V^{}=-1(-2)\%$. 
The black, red, green and blue curves show the results in the Type-I, -II, -X and -Y C2HDM, respectively, while 
the solid, dashed and dotted curves show the case for $f=\infty$, 2200 (1500) GeV and 1780 (1250) GeV, respectively, for the left (right) panel.  
Each dot on the curve denotes the prediction with $\tan\beta = 1$ to 10 with its interval of 1, and the dot at the left edge on each curve corresponds to $\tan\beta = 1$. 
The triangles represent the prediction with $\theta = 0$.  
}
\label{fig2}
\end{center}
\end{figure}

As it has been discussed in Ref.~\cite{finger}, 
the type of Yukawa interactions can be determined by looking at the correlation between $\Delta\kappa_E$ and $\Delta\kappa_D$ in the E2HDM, 
where $E$ and $D$ represent a charged lepton and a down-type quark, respectively. 
Now, let us discuss the correlation between $\Delta\kappa_E$ and $\Delta\kappa_D$ in the C2HDM. 

In Fig.~\ref{fig2},  we plot the prediction in the four types of the Yukawa interaction on the $\Delta\kappa_E^{}$ and
$\Delta\kappa_D^{}$ plane with the fixed value of 
$\Delta \kappa_V^{}$ being $-1\%$ (left panel) and $-2\%$ (right panel). 
The range of $\tan\beta$ is taken from 1 to 10. In these plots, the value of $\theta$ is determined by fixing $\Delta\kappa_V^{}$ and $f$ (or equivalently $\xi$). 
For each type of Yukawa interaction, 
we take $f=1780$ GeV (dotted curve), 2200 GeV (dashed curve)  and $\infty$ (solid curve) for the left panel, while 
$f=1250$ GeV (dotted curve), 1500 GeV (dashed curve)  and $\infty$ (solid curve) for the right panel. 
From this figure, we can extract  two important aspects: (i) the models with a different type of Yukawa interaction can be 
separated by looking at $\Delta\kappa_E$ and $\Delta\kappa_D$ and (ii) for a fixed value of $\Delta\kappa_V^{}$ and the type of Yukawa interaction, 
predicted regions on the $\Delta\kappa_E$-$\Delta\kappa_D$ plane can be different depending on the value of $f$. 
It is also shown that the magnitude of $\Delta\kappa_{E,D}^{}$ with a smaller value of $f$ tends to be small for a given value of $\tan\beta$ as compared 
to that with a larger $f$. 
As an extreme case $f=1740\,(1230)$ GeV, where the deviation $\Delta\kappa_V^{}=-1\%\,(-2\%)$ comes only via the non-zero $\xi$ (or equivalently the case with $s_\theta=0$), 
the prediction is given as a point indicated by the triangle, because the $\tan\beta$ dependence vanishes in this case. 
In addition, the predicted region with a fixed range of $\tan\beta$ shrinks when the value of $f$ is getting small, because the 
$\tan\beta$ dependent part of $\kappa_f^{}$ is proportional to $s_\theta$ as seen in Eq~(\ref{xihh}). 

Therefore, if a non-zero value of $\Delta\kappa_V^{}$ is measured at  collider experiments, 
we have an indirect evidence for a non-minimal Higgs sector, possibly belonging to a E2HDM or C2HDM.  
Furthermore, by looking at the pattern of the deviations in $\Delta\kappa_E$ and 
$\Delta\kappa_D$, we can discriminate between the four types of Yukawa interactions. 
In particular, if the Type-X or Type-Y Yukawa interaction is realised, the composite dynamics can also be 
extracted  from the different allowed regions of the predictions on the $\Delta\kappa_E$--$\Delta\kappa_D$ plane. 
For the Type-I and Type-II cases, a prediction with a non-zero value of $\xi$ corresponds to the case with a different value of $\tan\beta$ in the E2HDM, so that 
we need to use  other information, such as the decay properties of the extra Higgs bosons as we will discuss  below. 
We note that making use of information from $\Delta\kappa_U^{}$ with $U$ being an up-type quark is also helpful 
to extract the sign of $s_\theta$ as long as $\tan\beta$ and/or $\xi$ are not very large. 

\subsection{Decays of extra Higgs bosons}

Next, we discuss the decay properties of the extra Higgs bosons $H$, $A$ and $H^\pm$ in  both the E2HDM and C2HDM with the four types of  Yukawa interaction. 
In particular, we compare the BRs of the extra Higgs bosons in the two models with the same value of $\Delta\kappa_V^{}$. 
As examples, we consider the following three benchmark points giving $\Delta\kappa_V^{} = -2\%$:
\begin{align}
\text{BP1}: (s_\theta,\xi) = (-0.2,0),~~
\text{BP2}: (s_\theta,\xi) = (-0.1,0.03),~~
\text{BP3}: (s_\theta,\xi) = (0,0.04).  \label{bps}
\end{align}
BP1 corresponds to the E2HDM case, while BP2 and BP3 are two possible C2HDM cases, the latter corresponding to  zero-mixing angle.

\begin{figure}[t]
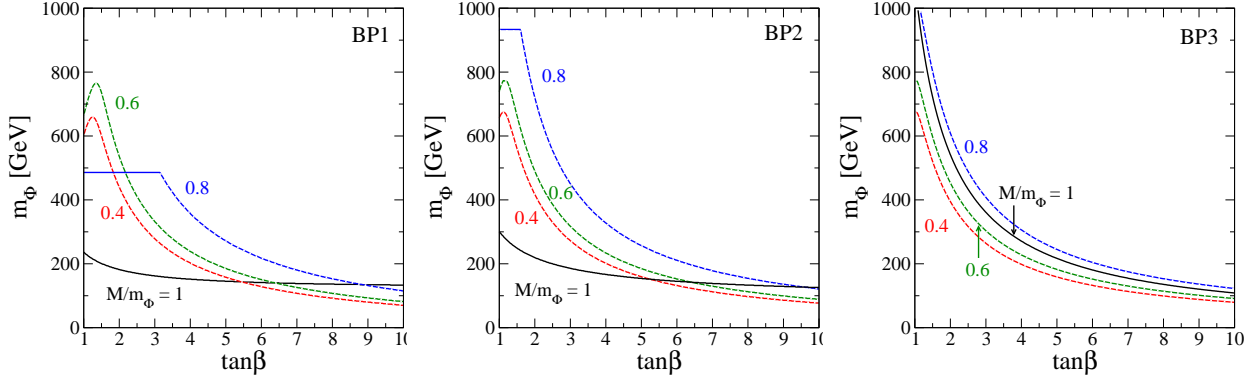

\begin{center}
\includegraphics[width=54mm]{uni_BP1.eps}\hspace{0mm}
\includegraphics[width=54mm]{uni_BP2.eps}\hspace{0mm}
\includegraphics[width=54mm]{uni_BP3.eps}
\caption{Upper limit on the mass parameter $m_\Phi^{}(=m_{H^\pm}^{}=m_A^{}=m_H^{})$ from the perturbative unitarity (indicated by the dashed curves) 
and the vacuum stability bounds (indicated by the solid curves) as a function of $\tan\beta$
in BP1 (left), BP2 (center) and BP3 (right). We take several fixed values of the ratio $M/m_\Phi^{}$ and $\sqrt{s}=1$ TeV for the unitarity bound. 
}
\label{fig:uni}
\end{center}
\end{figure}

Before studying the BRs, we survey the allowed parameter regions by bounds from the perturbative unitarity and the vacuum stability.  
Details of these bounds have been discussed in Ref.~\cite{DeCurtis:2016scv}. 
Concerning to the unitarity bound, we take into account all the elastic scatterings of 
2 body to 2 body scalar boson processes up to ${\cal O}(s^0)$ dependences, where $\sqrt{s}$ is the scattering energy.  
Differently from E2HDMs, the $s$-wave amplitude matrix has terms proportional to $s \, \xi$, thus indicating that an UV completion of the theory is needed at high energy. 
Here we fix $\sqrt{s}=1$ TeV.
 
\enlargethispage{15 mm}
In Fig.~\ref{fig:uni}, we show the allowed parameter region on the ($\tan\beta$, $m_{\Phi}^{})$ plane for the three benchmark points, where  $m_\Phi = m_{H^\pm} = m_A = m_H$.  
The region above each curve is excluded by perturbative unitarity or vacuum stability, so that this figure shows the absolute theoretical upper limit on $m_{\Phi}^{}$. 
Different colours of each curve show different choices of the ratio $M/m_\Phi^{}$ being 1, 0.8. 0.6 and 0.4. 
We can see that, typically, the unitarity and/or the vacuum stability bounds become stronger as the value of $\tan\beta$ increases. 
In addition, the case with $M/m_\Phi^{}\lesssim 1$ tends to have a larger allowed value of $m_{\Phi}^{}$ as compared to the case with $M/m_\Phi^{}=1$. 
Following this result, we take $\tan\beta = 2$, $m_{\Phi}^{}\leq 500$ GeV and $M/m_{\Phi}^{}=0.8$ for the following analysis. 

\begin{figure}[H]
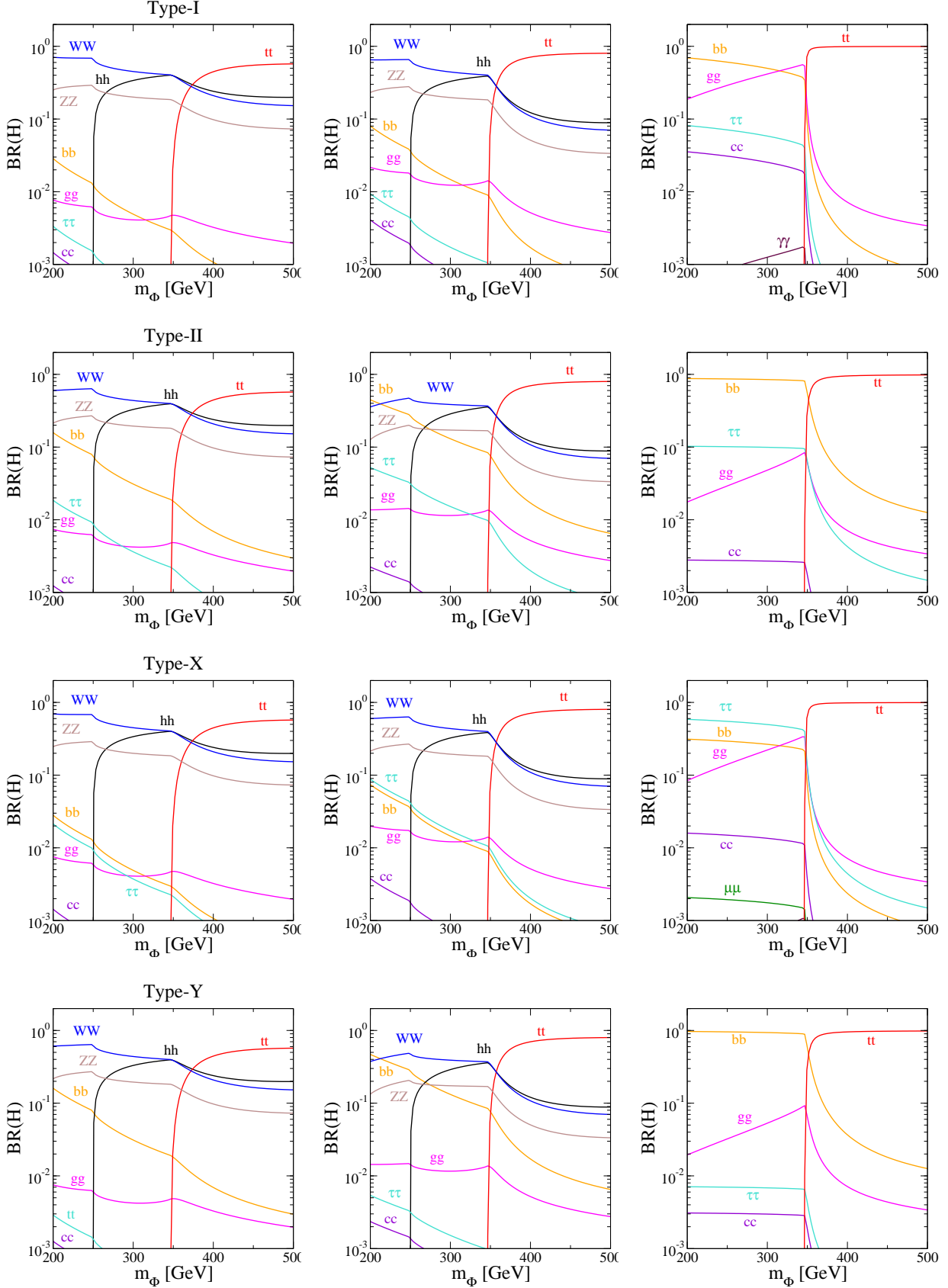

\begin{center}
\includegraphics[width=52mm]{BR_H_1_BP1.eps}\hspace{2mm}
\includegraphics[width=52mm]{BR_H_1_BP2.eps}\hspace{2mm}
\includegraphics[width=52mm]{BR_H_1_BP3.eps}\hspace{2mm}\\\vspace{4mm}
\includegraphics[width=52mm]{BR_H_2_BP1.eps}\hspace{2mm}
\includegraphics[width=52mm]{BR_H_2_BP2.eps}\hspace{2mm}
\includegraphics[width=52mm]{BR_H_2_BP3.eps}\hspace{2mm}\\\vspace{4mm}
\includegraphics[width=52mm]{BR_H_X_BP1.eps}\hspace{2mm}
\includegraphics[width=52mm]{BR_H_X_BP2.eps}\hspace{2mm}
\includegraphics[width=52mm]{BR_H_X_BP3.eps}\hspace{2mm}\\\vspace{4mm}
\includegraphics[width=52mm]{BR_H_Y_BP1.eps}\hspace{2mm}
\includegraphics[width=52mm]{BR_H_Y_BP2.eps}\hspace{2mm}
\includegraphics[width=52mm]{BR_H_Y_BP3.eps}
\caption{
BRs of $H$ as a function of $m_\Phi^{}(=m_H^{}=m_A^{}=m_{H^\pm})$ with $\tan\beta = 2$ and $M = 0.8\times m_\Phi^{}$
in the four types of Yukawa interaction. The left, center and right panels show the case for BP1, BP2 and BP3, respectively. }
\label{br_H2}
\end{center}
\end{figure}

\begin{figure}[H]
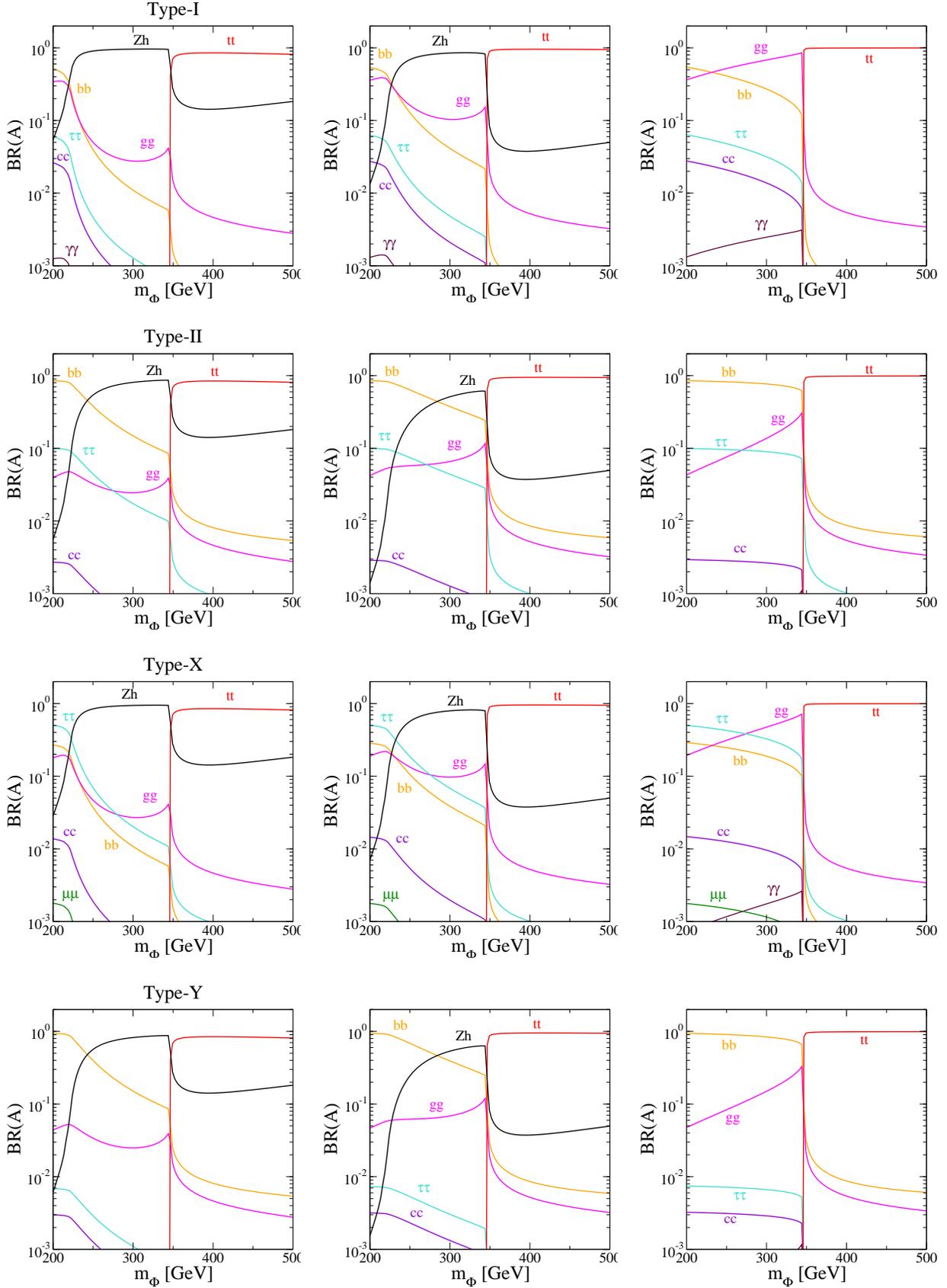

\begin{center}
\includegraphics[width=52mm]{BR_A_1_BP1.eps}\hspace{2mm}
\includegraphics[width=52mm]{BR_A_1_BP2.eps}\hspace{2mm}
\includegraphics[width=52mm]{BR_A_1_BP3.eps}\hspace{2mm}\\\vspace{4mm}
\includegraphics[width=52mm]{BR_A_2_BP1.eps}\hspace{2mm}
\includegraphics[width=52mm]{BR_A_2_BP2.eps}\hspace{2mm}
\includegraphics[width=52mm]{BR_A_2_BP3.eps}\hspace{2mm}\\\vspace{4mm}
\includegraphics[width=52mm]{BR_A_X_BP1.eps}\hspace{2mm}
\includegraphics[width=52mm]{BR_A_X_BP2.eps}\hspace{2mm}
\includegraphics[width=52mm]{BR_A_X_BP3.eps}\hspace{2mm}\\\vspace{4mm}
\includegraphics[width=52mm]{BR_A_Y_BP1.eps}\hspace{2mm}
\includegraphics[width=52mm]{BR_A_Y_BP2.eps}\hspace{2mm}
\includegraphics[width=52mm]{BR_A_Y_BP3.eps}
\caption{ Same as Fig.~\ref{br_H2}, but for the BRs of $A$.  \label{br_A2} }
\end{center}
\end{figure}

\begin{figure}[H]
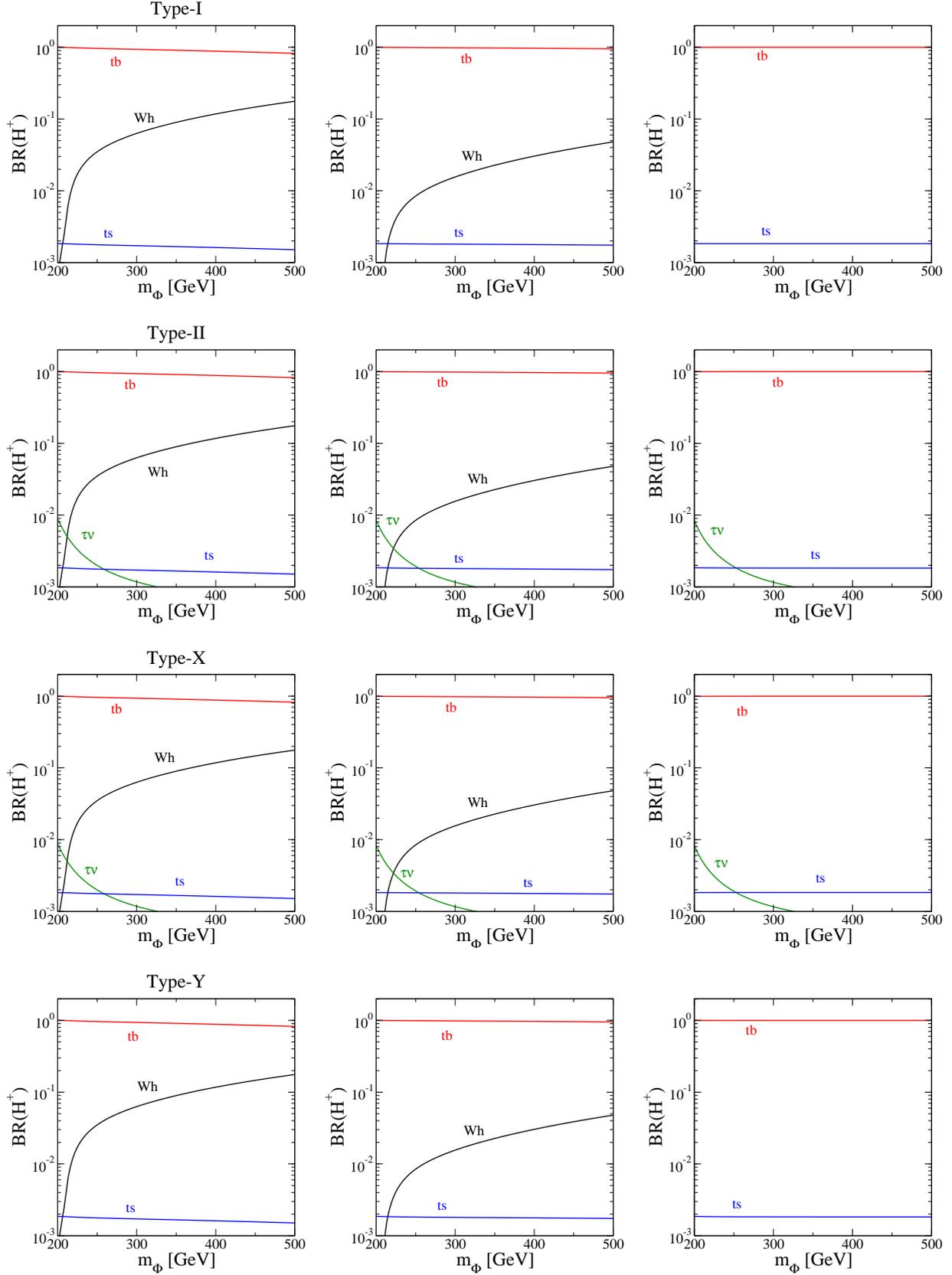

\begin{center}
\includegraphics[width=52mm]{BR_Hp_1_BP1.eps}\hspace{2mm}
\includegraphics[width=52mm]{BR_Hp_1_BP2.eps}\hspace{2mm}
\includegraphics[width=52mm]{BR_Hp_1_BP3.eps}\hspace{2mm}\\\vspace{4mm}
\includegraphics[width=52mm]{BR_Hp_2_BP1.eps}\hspace{2mm}
\includegraphics[width=52mm]{BR_Hp_2_BP2.eps}\hspace{2mm}
\includegraphics[width=52mm]{BR_Hp_2_BP3.eps}\hspace{2mm}\\\vspace{4mm}
\includegraphics[width=52mm]{BR_Hp_X_BP1.eps}\hspace{2mm}
\includegraphics[width=52mm]{BR_Hp_X_BP2.eps}\hspace{2mm}
\includegraphics[width=52mm]{BR_Hp_X_BP3.eps}\hspace{2mm}\\\vspace{4mm}
\includegraphics[width=52mm]{BR_Hp_Y_BP1.eps}\hspace{2mm}
\includegraphics[width=52mm]{BR_Hp_Y_BP2.eps}\hspace{2mm}
\includegraphics[width=52mm]{BR_Hp_Y_BP3.eps}
\caption{Same as Fig.~\ref{br_H2}, but for the BRs of $H^\pm$.\label{br_Hp2} }
\end{center}
\end{figure}

In Figs.~\ref{br_H2}, \ref{br_A2} and \ref{br_Hp2}, we respectively show the $m_\Phi^{}$ dependence of the BRs for $H$, $A$ and $H^\pm$ for  BP1 (left), BP2 (center) and BP3 (right) in the Type-I, -II, -X and -Y  configurations, respectively. 

When we look at the left and center panels of Fig.~\ref{br_H2}, we can observe the two thresholds at $m_H^{}\simeq 250$ GeV and 350 GeV which correspond to 
the $H\to hh$ and $H \to t\bar{t}$ channel, respectively. 
If we compare them and the right panels of Fig.~\ref{br_H2}, we find  significant differences in the $H$ decay modes. 
Namely, the $H\to VV$ ($V=W^+W^-,\,ZZ$) and $H \to hh$ modes are absent in the right panels, because they  are proportional to $s_\theta^2$. 
In addition, in the BP3 case, the difference among the four types of Yukawa interactions becomes more clear, because only the fermionic final states of the $H$ decay mode are dominant.

Concerning the BRs of $A$ (Fig.~\ref{br_A2}), it is seen that their behaviour is drastically changed at $m_A^{}\simeq 350$ GeV.  
Namely, below $m_A^{}\simeq 350$ GeV, the $A\to Z^{(*)}h$ channel can be dominant in BP1 and BP2 depending on $m_\Phi^{}$, while the $A\to b\bar{b}$, $\tau^+\tau^-$ and/or $gg$
modes are dominant in BP3 depending on the type of Yukawa interactions. 
Notice that we have taken into account the three body decay process  $A \to Z^*h \to f\bar{f}h $, which becomes important when $m_A^{}< m_Z^{}+m_h\simeq 215$ GeV.  
Conversely, above $m_A^{}\simeq 350$ GeV, $A\to t\bar{t}$ becomes dominant in all four type models and all three benchmark points. 
In BP1 (BP2) with $m_A^{}\simeq 350$ GeV, $A \to Z^{(*)}h$ can be $10$--$20\%$ (a few \%) level depending on $m_\Phi^{}$ and the type of Yukawa interactions. 
Regarding the results for BP3, the behaviour of the BRs of $A$ is almost the same as those of $H$, where 
the $A\to Z^{(*)}h$ mode does not appear, because its decay rate is proportional to $s_\theta^2$. 
Only for the BR of the $A\to gg$ mode, it is slightly larger than that of the $H\to gg$ mode when we compare them 
with the same configuration, because of the difference in the loop function.  

The mass dependence on the BRs of $H^\pm$ is shown in Fig.~\ref{br_Hp2}. 
We see that the $H^+\to t\bar{b}$ mode is dominant in all the four types of Yukawa interactions and all the three benchmark points. 
In BP1 (BP2), the $H^+\to W^{+(*)}h$ mode can be about 20\% (5\%) at $m_{\Phi}^{}\simeq 500$ GeV.

\subsection{Productions of extra Higgs bosons at the LHC}

Finally, we discuss the production cross sections of the extra Higgs bosons at the LHC. 
We here consider the gluon fusion process $gg\to H/A$, the bottom quark associated process $gg \to b\bar{b}H/A$ and 
the gluon-bottom fusion process $g\bar{b} \to H^+ \bar{t}$. 
In fact the cross sections for the vector boson fusion ($qq' \to qq'H$) and vector boson associated ($q\bar{q}\to ZH$ and $q\bar{q}'\to W^\pm H^\mp$) processes 
are negligibly small, because of the suppressed gauge-gauge-Higgs couplings (by $s_\theta$). 

For the calculation of the gluon fusion cross section, we use the following equation:
\begin{align}
\sigma(gg\to \phi^0) = \frac{\Gamma(\phi^0 \to gg)}{\Gamma(h_{\text{SM}}\to gg)}\times \sigma (gg \to h_{\text{SM}}),~~(\phi^0=H~\text{or}~A), 
\end{align}
where $h_{\text{SM}}$ is the SM Higgs boson with the mass artificially set at $m_{\phi^0}$. 
We adopt the value of the gluon fusion cross section $\sigma (gg \to h_{\text{SM}})$ in the SM from Ref.~\cite{higgs}. 
%
For the other calculations of  the production cross sections, we use {\tt CalcHEP}~\cite{calchep} and adopt the 
{\tt CTEQ6L}~\cite{cteq} for the parton distribution functions with factorisation/renormalisation scale set at $Q=\sqrt{\hat s}$. 
We note that  the lepton Yukawa coupling  is not relevant for the calculation of the production cross sections, so that the 
result in the Type-I (Type-II) and Type-X (Type-Y) models are the same with each other. 
As in the previous subsection, we take BP1, BP2 and BP3 given in Eq.~(\ref{bps}) and $\tan\beta = 2$ for the numerical analysis. 

\begin{figure}[h]
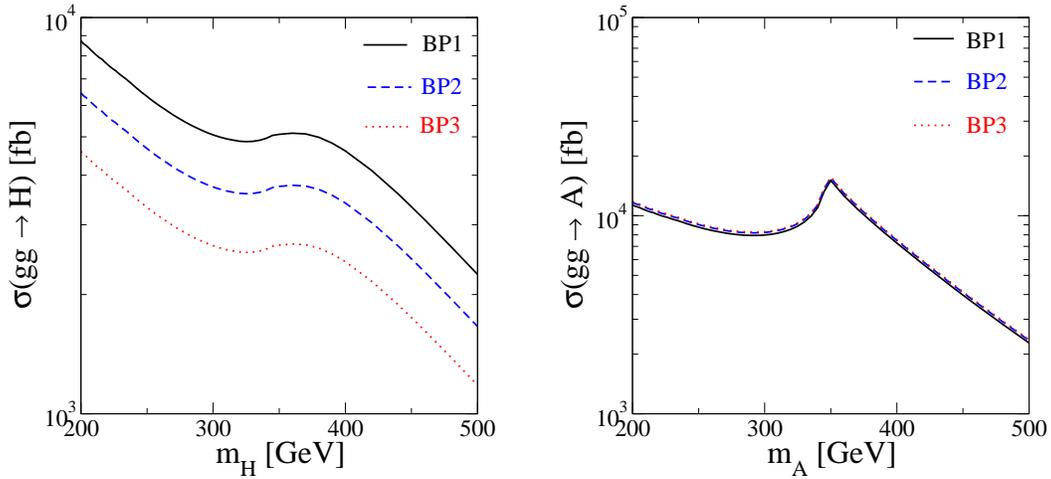

\begin{center}
\includegraphics[width=65mm]{gfusion_H_1_tb2.eps}\hspace{7mm}
\includegraphics[width=65mm]{gfusion_A_1_tb2.eps}
\caption{
Cross section of the gluon fusion process for $H$ (left) and $A$ (right) as a function of the extra neutral Higgs boson mass at $\sqrt{s}=13$ TeV 
in BP1, BP2 and BP3 with $\tan\beta = 2$. 
}
\label{gf}
\end{center}
\end{figure}

In Fig.~\ref{gf}, we show the gluon fusion production cross section as a function of the mass of the produced Higgs boson. 
In this process, the dependence of the type of Yukawa interactions is almost negligible, because only the top Yukawa coupling is important to determine the 
size of the cross section. 
The results for BP1, BP2 and BP3 are respectively shown as the solid, dashed, and dotted curves. 
We find differences in the cross section of $gg \to H$ among the three benchmark points, which   
comes from the $s_\theta$ term in $\zeta_H$ or $\xi_H$ given in Eq.~(\ref{xibh}). 
In contrast, the cross section for $A$ is essentially the same for the three benchmark points. 

\begin{figure}[h]
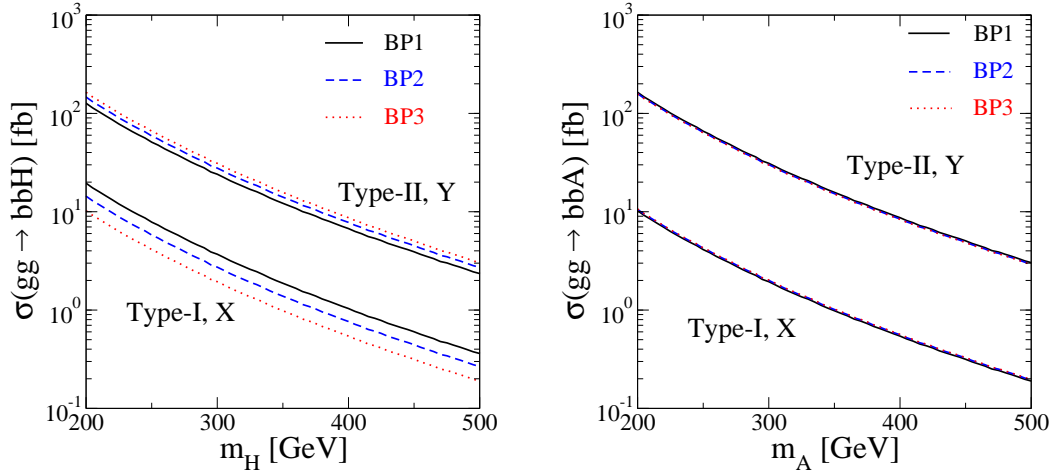

\begin{center}
\includegraphics[width=65mm]{bbH_1_tb2.eps}\hspace{7mm}
\includegraphics[width=65mm]{bbA_1_tb2.eps}
\caption{
Cross section for the bottom quark associated production process  for $H$ (left) and $A$ (right) as a function of the extra neutral Higgs boson mass at $\sqrt{s}=13$ TeV 
in BP1, BP2 and BP3 with $\tan\beta = 2$. }
\label{ba}
\end{center}
\end{figure}

In Fig.~\ref{ba}, we show the cross section of the bottom quark associated production as a function of the mass of the produced Higgs boson. 
Typically, the cross section is more than one order of magnitude smaller than  
the gluon fusion production process because of the smallness of the bottom Yukawa coupling and the three body phase space. 
Differently from the gluon fusion, the dependence of the type is important, because the bottom Yukawa coupling determines 
the size of the cross section. 
In fact, the cross section in Type-II and Type-Y is almost one order of magnitude greater than that in Type-I and Type-X. 
Similar to the case for the gluon fusion, a larger discrepancy of the cross section among BP1, BP2 and BP3 
is seen for the production of $H$ state, as for the $A$ one differences are marginal. 

\begin{figure}[h]
\begin{center}
\includegraphics[width=65mm]{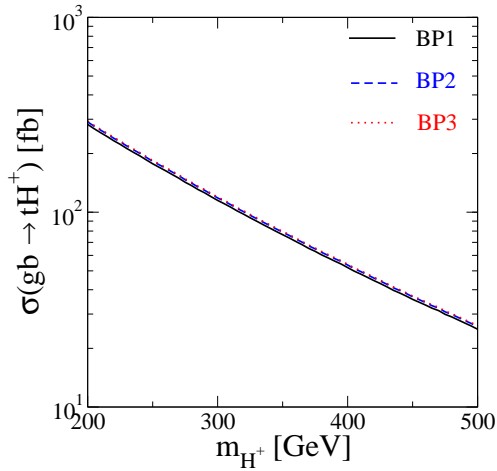}
\caption{
Cross section of the gluon bottom fusion process of $H^\pm$ as a function of the extra neutral Higgs boson mass at $\sqrt{s}=13$ TeV 
in BP1, BP2 and BP3 with $\tan\beta = 2$. 
}
\label{gb}
\end{center}
\end{figure}

In Fig.~\ref{gb}, we show the gluon-bottom fusion production cross section for the $H^\pm$ state. 
Similar to the gluon fusion process, the dependence of the type of Yukawa interactions is almost negligible, because of involving the top Yukawa coupling. 
The differences among the three benchmark points are negligibly small. 

To summarise, in this section, we have discussed the differences between the E2HDM and C2HDM 
by focusing on the deviations in the SM-like Higgs boson couplings from the SM predictions as well as the decay BRs and production cross sections at the LHC. 
We have shown that, even if both the E2HDM and C2HDM give the same value of the deviation in the $hVV$ coupling, 
we can find significant differences in the correlation of $\Delta\kappa_E^{}$-$\Delta \kappa_D^{}$ in the two scenarios
(elementary and composite). 
In addition, through the combination of the differences in the decay 
BRs and production cross sections for the extra Higgs bosons, we may be able to distinguish these two hypothesis on the nature of the Higgs bosons responsible for EWSB.

\section{Conclusions}

In this paper, we have continued our exploration of C2HDM scenarios, started with Ref.~\cite{DeCurtis:2016scv}, assuming 
four different types of Yukawa interactions, wherein the nature of all Higgs states is such that they are composite objects. Specifically, they are the pNGBs from the global
symmetry breaking $SO(6)\to  SO(4)\times SO(2)$, induced explicitly by interactions between a new strong sector and the SM fields at the compositeness scale $f$. Such pNGBs, for which we adopt the same scalar potential as in the E2HDM, then trigger EWSB governed by the SM gauge group. Under the assumption of partial compositeness, it is rather    natural that
one of the emerging physical Higgs fields, the lightest one, is the 125 GeV state, $h$,   discovered at CERN.  

Within this construct, we then proceed to carry out a phenomenological study aiming at establishing the potential of the 
LHC in
disentangling the two hypotheses, E2HDM versus C2HDM, by exploiting the fact that drastically different production and decay patterns for  the four heavy Higgs states ($H, A$ and $H^\pm$) 
may onset in the composite scenario with respect to the elementary one, even when the properties of 
SM-like Higgs state are the same (within experimental accuracy) in the two scenarios. This has been done after imposing both theoretical (already derived in Ref. \cite{DeCurtis:2016scv}) and experimental (obtained here by suitably modifying numerical toolboxes 
used in E2HDM analysis to also embed the C2HDM option) constraints,
the latter revealing a marked dependence upon $\xi$ only for the case of Type-I Yukawa interactions. Specifically, the most dramatic situation could occur when, e.g.,
in the presence of an established deviation of a few percents from the SM prediction for the $hVV$  ($V=W^\pm,Z$)
 coupling (in fact, possibly the most precisely determined one at the LHC), the E2HDM would require
the mixing between the $h$ and $H$ states to be non-zero whereas in the C2HDM compliance with such a measurement could be achieved also for the zero mixing case. Hence, in this situation, the $H\to W^+W^-$ and $ZZ$ decays would be forbidden in the composite case, while still being allowed in the elementary one. (Similarly, Higgs-strahlung and vector-boson-fusion would be nullified in the C2DHM scenario, unlike in the E2HDM, while potentially large differences  also appear in the case
of gluon-gluon fusion and associated production with $b\bar b$ pairs.) Clearly, also intermediate situations can be
realised. Therefore, a close scrutiny of the possible signatures of a heavy CP-even Higgs boson, $H$, would be a key to assess the viability of either model. Regarding the CP-odd Higgs state, $A$, in the extreme case of non-zero(zero)
mixing in the E2HDM(C2HDM), again, it is the absence of a decay, i.e., $A\to Zh$,  in the C2HDM that would distinguish it from the E2HDM.         
In the case of the $H^\pm$ state, a similar role is played by the $H^\pm\to W^\pm h$ decay. Obviously, for  both these states too, intermediate situations are also possible, so that a precise study of these two channels would be a 
further strong handle to use in order to disentangle the two hypotheses. As far as $A$ and $H^\pm$ production modes which are accessible at the
LHC, i.e., gluon-gluon fusion and associate production with $b\bar b$  pairs (for the $A$) 
and associated production with $b\bar t$ pairs (for the $H^+$), 
are concerned though, practically no difference appears. The actual size of all these differences between the E2HDM and C2HDM
is  governed by the
value of the $\xi= v_{\rm SM}^2/f^2$ parameter, the larger the latter the more significant the former. Finally, although
there are quantitative differences between the usual four Yukawa types (I, II, X and Y, in our notation) when predicting the yield of both the E2HDM and C2HDM, the qualitative pattern we described would generally persist. In fact, a similar phenomenology would emerge if deviations were instead (or in addition) established in the Yukawa couplings of the $h$ state to $b$-quarks and/or $\tau$-lepton.  

In short, if deviations will be established during Run 2 of the LHC in the couplings of the discovered Higgs state with 
either SM  gauge bosons or matter fermions, then, not only a thorough investigation of the 2HDM hypothesis is called for (as one of the simplest non-minimal version of EWSB induced by the Higgs mechanism via {\sl doublet} states, like the one already
discovered) but a dedicated scrutiny of the decay patters of all potentially accessible heavy Higgs states could enable one
to separate the E2HDM from the C2HDM.

\noindent
\section*{Acknowledgments}
\noindent 

The work of  SM is financed in part through the NExT Institute and by the STFC Consolidated Grant ST/J000391/1.
This work was supported by a JSPS postdoctoral fellowships for research abroad (KY).
EY was supported by the Ministry of  National Education of Turkey.

\begin{appendix}

\section{Feynman rules}

\begin{table}[t]
\begin{center}
{\renewcommand\arraystretch{1}
\begin{tabular}{c|cc}\hline\hline
Vertex & Coefficient    \\\hline 
$H^\pm \overleftrightarrow{\partial}_\mu  A W^{\mp \mu}$  &  $\frac{g}{2}$             \\\hline 
$H^\pm \partial_\mu  h W^{\mp \mu}$  &  $\mp i\frac{g}{2}(1-\frac{5}{6}\xi)\sin\theta$             \\\hline 
$h \partial_\mu  H^\pm W^{\mp \mu}$  &  $\pm i\frac{g}{2}(1-\frac{1}{6}\xi)\sin\theta$             \\\hline 
$H^\pm \partial_\mu  H W^{\mp \mu}$  &  $\mp i\frac{g}{2}(1-\frac{5}{6}\xi)\cos\theta$             \\\hline 
$H \partial_\mu  H^\pm W^{\mp \mu}$  &  $\pm i\frac{g}{2}(1-\frac{1}{6}\xi)\cos\theta$             \\\hline 
$A \partial_\mu  h Z^\mu$  &  $  -\frac{g_Z}{2}(1-\frac{5}{6}\xi)\sin\theta$             \\\hline 
$h \partial_\mu  A Z^{\mu}$  &  $ \frac{g_Z}{2}(1-\frac{1}{6}\xi)\sin\theta$             \\\hline 
$A \partial_\mu  H Z^{\mu}$  &  $ -\frac{g_Z}{2}(1-\frac{5}{6}\xi)\cos\theta$             \\\hline 
$H \partial_\mu  A Z^{\mu}$  &  $ \frac{g_Z}{2}(1-\frac{1}{6}\xi)\cos\theta$             \\\hline 
$H^+ \overleftrightarrow{\partial}_\mu  H^- Z^{\mu}$  &  $-i\frac{g_Z}{2} c_{2W}$             \\\hline
$H^+ \overleftrightarrow{\partial}_\mu  H^- A^{\mu}$  &  $-ie$             \\\hline\hline 
\end{tabular}}
\caption{Coefficients of the Scalar-Scalar-Gauge type vertices. }
\label{FR1}
\end{center}
\end{table}

We present the trilinear couplings of the Higgs bosons which are relevant to the discussion of the phenomenology given in Sec.~III.  
First, the Gauge-Gauge-Scalar type interactions are given by 
\begin{align}
{\cal L}_{\text{kin}} &=  \left(1-\frac{\xi}{2}\right)(h\cos\theta-H\sin\theta)\left(gm_W W_\mu^+W^{-\mu} + \frac{g_Z^{}}{2}m_ZZ_\mu Z^\mu\right). 
\end{align}
Second, the coefficients of the Scalar-Scalar-Gauge type interactions are extracted as given in Table~\ref{FR1}, where 
we introduce 
\begin{align}
X\overleftrightarrow{\partial}_\mu Y = X(\partial _\mu Y) - (\partial _\mu X) Y. 
\end{align}
Finally, the scalar trilinear $Hhh$ and $H^+H^-h$ couplings defined by 
\begin{align}
{\cal L} = +\lambda_{Hhh}Hhh + \lambda_{H^+H^-h}H^+H^-h + \cdots 
\end{align}
are extracted by 
\begin{align}
%
\lambda_{Hhh} & = \frac{s_\theta}{v_{\text{SM}}s_{2\beta}}\left[  
-\frac{s_{2(\beta+\theta)}}{2}(2m_h^2 + m_H^2) + \frac{1}{2}(s_{2\beta}+3s_{2(\beta+\theta)})M^2\right] \notag\\
& + \frac{\xi}{12v_{\text{SM}}}s_\theta \left[m_H^2-2m_h^2 + (1+3c_{2\theta}+6\cot 2\beta s_{2\theta})M^2\right], \\
\lambda_{H^+H^-h} & = \frac{c_\theta}{v_{\text{SM}}}\left[-(1+2\cot2\beta\tan\theta)m_h^2-2m_{H^\pm}^2+\frac{2}{s_{2\beta}c_\theta}s_{2\beta +\theta }M^2\right]\notag\\
&+\frac{\xi}{6v_{\text{SM}}}c_{\theta}\left[(1+4\cot2\beta\tan\theta)m_h^2+2(m_{H^\pm}^2-M^2) \right]. 
\end{align}
\end{appendix}

\enlargethispage{28.5mm}

\end{document}